\documentclass[aps,prd,reprint,twocolumn,superscriptaddress,longbibliography,nofootinbib,floatfix,showpacs]{revtex4-1}
\usepackage{epsfig}
\usepackage{amsmath,amssymb,amsfonts}
\usepackage{hyperref}
\usepackage{mathrsfs}
\usepackage{bbm}
\usepackage{slashed}
\usepackage{graphicx}
\usepackage{verbatim}

\usepackage{bm}

\usepackage[usenames]{color}

\usepackage{tikz}
\usetikzlibrary{calc,decorations.markings,decorations.pathmorphing,positioning,shapes,snakes,arrows}

\definecolor{darkgreen}{rgb}{0.2,0.6,0}

\newcommand{\be}{\begin{equation}}
\newcommand{\ee}{\end{equation}}
\newcommand{\bw}{\begin{widetext}}
\newcommand{\ew}{\end{widetext}}
\newcommand{\bi}{\begin{itemize}}
\newcommand{\ei}{\end{itemize}}
\newcommand{\bea}{\begin{eqnarray}}
\newcommand{\eea}{\end{eqnarray}}
\newcommand{\bracket}[2]{\bra{#1}\,#2\rangle} 
\newcommand{\bra}[1]{\langle\,#1\,|}          
\newcommand{\ket}[1]{|\,#1\,\rangle}          
\newcommand{\ud}{\mathrm{d}}

\newcommand{\LCm}{{\scriptscriptstyle -}} 
\newcommand{\LCp}{{\scriptscriptstyle +}}
\newcommand{\LCpm}{{\scriptscriptstyle \pm}}
\newcommand{\LCmp}{{\scriptscriptstyle \mp}}

\newcommand{\LCperp}{{\scriptscriptstyle \perp}}

\usepackage[T1]{fontenc} \usepackage[latin1]{inputenc}

\begin{document}

\title{Nonlinear photon trident versus double Compton scattering and resummation of one-step terms}

\author{Greger Torgrimsson}
\email{g.torgrimsson@hzdr.de}
\affiliation{Helmholtz-Zentrum Dresden-Rossendorf, Bautzner Landstra{\ss}e 400, 01328 Dresden, Germany}
\affiliation{Theoretisch-Physikalisches Institut, Abbe Center of Photonics,
	Friedrich-Schiller-Universit\"at Jena, Max-Wien-Platz 1, D-07743 Jena, Germany}
\affiliation{Helmholtz Institute Jena, Fr\"obelstieg 3, D-07743 Jena, Germany}

\begin{abstract}

We study the photon trident process, where an initial photon turns into an electron-positron pair and a final photon under a nonlinear interaction with a strong plane-wave background field. We show that this process is very similar to double Compton scattering, where an electron interacts with the background field and emits two photons. 
We also show how the one-step terms can be obtained by resumming the small- and large-$\chi$ expansions. We consider a couple of different resummation methods, and also propose new resummations (involving Meijer-G functions) which have the correct type of expansions at both small and large $\chi$. These new resummations require relatively few terms to give good precision. 
	
\end{abstract}	

\maketitle

\section{Introduction}

The strength of a high-intensity laser is usually expressed in terms of $a_0=E/\omega$\footnote{We use units with $m_e=c=\hbar=1$ and absorb a factor of $e$ into the background field, $eE\to E$. We use $g_{\mu\nu}={\rm diagonal}(1,-1,-1,-1)$.}, where $E$ is the field strength and $\omega$ a typical frequency scale. The $\mathcal{O}(\alpha)$ processes nonlinear Compton scattering $e^\LCm\to e^\LCm+\gamma$ followed by nonlinear Breit-Wheeler pair production $\gamma\to e^\LCm+e^\LCp$ were observed more than two decades ago at SLAC~\cite{Bamber:1999zt}. There the lasers had $a_0<1$ and the observation could be explained in terms of perturbative (albeit mutliphoton/nonlinear) physics. Today's lasers can have much larger $a_0$. For sufficiently large $a_0$ (depending on the size of the other parameters of the system), one can approximate $\mathcal{O}(\alpha^{n\geq2})$ processes by incoherent products of sequences of $\mathcal{O}(\alpha)$ processes, where the laser can be approximated as locally constant during each $\mathcal{O}(\alpha)$ step.  
This is a key ingredient of particle-in-cell codes~\cite{RidgersCode,Gonoskov:2014mda,Osiris,Smilei}, which are often the only means available to study higher-order processes. Since higher-order processes are expected to be important in upcoming high-intensity laser experiments, and since they are in general too difficult to compute exactly, it is important to 
\bi
\item[1)] study how to approximate $\mathcal{O}(\alpha^{n\geq2})$ processes, and
\item[2)] to more precisely estimate the size of the corrections and to delineate the region where these corrections can be neglected. 
\ei
1) involves for example the question how to sum over the spin and polarization of intermediate particles~\cite{Ritus:1972nf,Baier,King:2013osa,Morozov:1975uah}, which we have recently treated with Stokes vectors and ``strong-field-QED Mueller matrices'' in~\cite{Dinu:2019pau}.
For 2) it is natural to study in detail the $\mathcal{O}(\alpha^2)$ processes, for which one can with some effort calculate the entire probability. The trident process ($e^\LCm\to2e^\LCm+e^\LCp$) has been studied in~\cite{Ritus:1972nf,Baier,King:2013osa,Ilderton:2010wr,Hu:2010ye,Dinu:2017uoj,King:2018ibi,Mackenroth:2018smh,Torgrimsson:2020wlz} and double Compton scattering ($e^\LCm\to e^\LCm+2\gamma$) has been studied in~\cite{Morozov:1975uah,UnruhDoubleCompton,Lotstedt:2009zz,Loetstedt:2009zz,Seipt:2012tn,Mackenroth:2012rb,King:2014wfa,Dinu:2018efz}. Both these processes have a single particle (apart from the laser field) in the initial state. Processes with two particles in the initial state have recently attracted more interest~\cite{Bragin:2020akq,Blackburn:2020fqo,Tang:2019ffe,Ilderton:2019bop}, but they are quite different from both a conceptual and a calculational point of view.

However, there is one $\mathcal{O}(\alpha^2)$ process that has not received much attention, namely the photon trident process ($\gamma\to e^\LCm+e^\LCp+\gamma$), which is, like trident and double Compton, also a $\mathcal{O}(\alpha^2)$ process with only one initial particle. To the best of our knowledge, this process has only been studied in~\cite{MorozovNarozhnyiPhTr}. So, our goal in this paper is to study this process. 

We use the same methods as we previously used in~\cite{Dinu:2017uoj,Dinu:2018efz} to study the trident and double Compton. Although those two processes had already been studied in a couple papers, we were able to show that certain terms that had been omitted in the previous literature on the locally-constant-field (LCF) regime are actually crucial for point 2). In fact, for double Compton we showed in~\cite{Dinu:2018efz} that the inclusion of the omitted terms can even change the order of magnitude of the correction.
These terms are part of what we call the one-step part of the probability, which gives the correction to the two-step part, i.e. the incoherent product of two $\mathcal{O}(\alpha)$ processes summed over the spin/polarization of the intermediate particle. In the LCF regime one can expand the probability in a power series in $1/a_0\ll1$. The two-step scales as $\mathbb{P}_2\sim a_0^2+\mathcal{O}(a_0^0)$ and the one-step as $\mathbb{P}_1\sim a_0$. The terms that were omitted in the previous LCF literature are the exchange part of the probability, by which we mean the cross-term between the two parts of the amplitude that are related by swapping place of the two identical particles in the final state. We call the non-exchange part of the probability the direct part\footnote{So, ``direct'' $\ne$ ``one-step''. Instead, $\mathbb{P}_2=\mathbb{P}_2^{\rm dir}$ and $\mathbb{P}_1=\mathbb{P}_1^{\rm dir}+\mathbb{P}_1^{\rm ex}$.}.
In~\cite{Dinu:2018efz} we showed that the most difficult part of the exchange term in double Compton has the same functional form as the corresponding term in trident, so one can obtain one from the other by some simple replacement of the parameters. Since these terms are the most difficult to calculate, this close relation is of course very useful in practice, as it mean that we can calculate them using the same methods.

In this paper we show that photon trident has an even closer relation to double Compton, as expected. Indeed, on an analytical level, all contributions to photon trident can be obtained from the corresponding terms in double Compton by a simple replacement of the longitudinal momenta. For $\chi\ll1$ we show explicitly that it is possible to obtain the spectrum by replacing the longitudinal momenta in the double-Compton spectrum. Here $\chi=a_0b_0$, where $b_0=kp$ is the product of the wave vector of the laser ($k_0=\omega$) and the momentum $p_\mu$ of the initial particle.
In the photon-trident case there are no identical particles in the final state, but there are two different contributions to the amplitude where the final photon is emitted by either the electron or the positron, and the cross-term between those diagrams corresponds to the exchange terms in trident and double Compton. Our results thus show that the most complicated terms in all these three second-order processes are closely related.

However, since these replacements involve e.g. changing sign of some lightfront-longitudinal momenta (which are all positive for real particles), these relations cannot be used to simply directly translate numerical results of e.g. the spectrum in double Compton into results for photon trident. In particular, in~\cite{Dinu:2018efz} we showed that the direct and exchange parts of the one-step tend to cancel, but from these relations alone we cannot say whether this cancellation also happens in photon trident. To answer this question we have to perform new calculations.

While the one-step can be computed numerically as in~\cite{Dinu:2018efz}, here we will show that another way is to use resummation methods. One can consider expansions in different parameters. Here we will consider the small- and large-$\chi$ expansions. We will show that existing resummation methods based on Borel transformation, conformal maps and Pad\'e approximants can be used. In some cases it can become time consuming to calculate many orders in these expansions, so we want resummation methods that maximize the precision over larger $\chi$ intervals given a finite number of terms. New resummation methods such as the one in~\cite{Alvarez:2017sza}, can be used to improve the resummation. However, with these general resummation methods there is still room for improvement. So, we have found new resummation methods which are tailor-made for strong-field QED in LCF. These new resummations have the same type of expansions as the exact result for both small and large $\chi$. This means that we need relatively few terms from these expansions in order to find precise resummations over large intervals of $\chi$. In fact, this allows us to find uniform resummations that works for any value of $\chi$. 

This paper is organized as follows. In Sec.~\ref{Definitions and derivation section} we give the necessary definitions and explain how to derive the exact results for photon trident. The exact results are presented and compared with double Compton in Sec.~\ref{Exact results section}. In Sec.~\ref{Saddle point approximations section} we derive saddle-point approximations to further compare with double Compton. In Sec.~\ref{resumSmallSection} we show how to resum the small-$\chi$ expansion. In Sec.~\ref{resumLargeSection} we derive the large-$\chi$ expansion and present a new resummation in terms of a sum over Meijer-G functions. In Sec.~\ref{resumSmallLargeSection} we present another new resummation, which is a sum of terms that are quadratic in Meijer-G functions and which we show can be used to resum the small- and large-$\chi$ expansions simultaneously. Having used double Compton in Sec.~\ref{resumSmallSection}, \ref{resumLargeSection} and~\ref{resumSmallLargeSection} as an example for these resummation approaches, in Sec.~\ref{photonTridentResum} we use them for photon trident. We also present new resummations which involve sums of products of Airy functions.
We conclude in Sec.~\ref{conclusionsSection}.

\section{Definitions and derivation}\label{Definitions and derivation section}

We consider in general pulsed plane-wave background fields. The structure of this field makes it useful to use lightfront coordinates $v^\LCpm=2v_\LCmp=v^0\pm v^3$, $v^\LCperp=\{v^1,v^2\}$. For momentum variables we use $\bar{P}=\{P_\LCm,P_\LCperp\}$. The field depends only on one lightlike coordinate, which is chosen to be $x^\LCp$ and is referred to as lightfront time. Instead of $x^\LCp$ we usually use $\phi=kx=\omega x^\LCp$ as integration variable, where $\omega$ is a characteristic frequency of the field. 
In terms of these coordinates the field can be expressed as $f_{\mu\nu}=k_\mu a'_\nu-k_\nu a'_\mu$, where $a_\LCp=a_\LCm=0$ and $a_\LCperp(\phi)$ has an arbitrary pulse shape and arbitrary polarization.

The initial state contains a photon with momentum $l_\mu$ and polarization $\varepsilon_\mu$. The photon is on-shell so $l_\LCp=l_\LCperp^2/(4l_\LCm)$. We use lightfront gauge where $k\varepsilon=0$, so $\varepsilon_\LCm=0$ and $\varepsilon_\LCp=l_\LCperp\varepsilon_\LCperp/(2l_\LCm)$. Although we do not consider any nontrivial wave-packet effects here, it is still convenient to start with an initial state described by a wave-packet $f(l)$ as
\be
\ket{{\rm in}}=\int\!\ud\tilde{l}\;f(l)\varepsilon^\mu \hat{a}_\mu^\dagger(l)\ket{0} \;,
\ee
where the momentum measure 
\be
\ud\tilde{P}=\frac{\ud^2 P_\LCperp\ud P_\LCm\theta(P_\LCm)}{(2\pi)^3 2P_\LCm}
\ee
is Lorentz-invariant. The step function comes from the fact that the longitudinal momentum $P_\LCm=P_0-P_3>0$ for all physical momenta.
The photon mode operator obeys
\be
[\hat{a}_\mu(l),\hat{a}_\nu^\dagger(l')]=-2l_\LCm\bar{\delta}(l-l')L_{\mu\nu} \;,
\ee
where $\bar{\delta}(P)=(2\pi)^3\delta(P_\LCm)\delta^2(P_\LCperp)$ and
\be
L_{\mu\nu}(l)=g_{\mu\nu}-\frac{k_\mu l_\nu+l_\mu k_\nu}{kl} \;.
\ee
The sum over two orthogonal polarization vectors, e.g. with $\varepsilon_\LCperp=\{1,0\}$ and $\varepsilon_\LCperp=\{0,1\}$, is given by
\be
\sum_{\rm pol.}\epsilon_\mu(l)\epsilon_\nu(l)=-L_{\mu\nu}(l) \;.
\ee 
We assume that the wave-packet is sharply peaked, so ($\bracket{{\rm in}}{{\rm in}}=1$)
\be
\int\ud\tilde{l}|f(l)|^2 F(l)=F(l) \;,
\ee
where we also use $l$ for the position of the peak.

We are interested in the probability that this initial state decays into a final state with an electron, a positron and a photon with momentum $p_\mu$, $p'_\mu$ and $l'_\mu$, respectively. The amplitude for this process has two terms ($M=M_e+M_p$), where the final photon is emitted by either the electron ($M_e$) or the positron ($M_p$). 
\begin{figure}
\includegraphics[width=\linewidth]{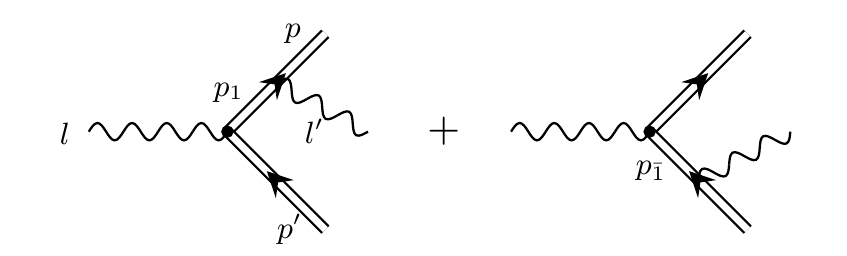}
\caption{Feynman diagrams for photon trident.}
\label{FeynmanDiagramFig}
\end{figure}
These are given by, see Fig.~\ref{FeynmanDiagramFig},
\be
\begin{split}
&\frac{1}{k_\LCp}\bar{\delta}(p+p'+l'-l)M_e= \\
&(-ie)^2\int\ud^4x_1\ud^4x_2\bar{\psi}(x_2)\slashed{\varepsilon}'e^{il'x_2}S(x_2,x_1)\slashed{\varepsilon}e^{-ilx_1}\psi_\LCm(x_1)
\end{split}
\ee
and
\be
\begin{split}
&\frac{1}{k_\LCp}\bar{\delta}(p+p'+l'-l)M_p= \\
&(-ie)^2\int\ud^4x_1\ud^4x_2\bar{\psi}(x_1)\slashed{\varepsilon}e^{-ilx_1}S(x_1,x_2)\slashed{\varepsilon}'e^{il'x_2}\psi_\LCm(x_2) \;,
\end{split}
\ee
where $\psi(p,x)=K(p,\phi)u(p,\sigma)\varphi(p,x)$ is the Volkov solution, where
\be
\varphi(p,x)=\exp\left\{-i\left(px+\int^{kx}\ud\phi\frac{2ap-a^2}{2kp}\right)\right\} \;,
\ee
\be
K(p,\phi)=1+\frac{\slashed{k}\slashed{a}}{2kp}
\ee
and $u(p,\sigma_p)$ is a field-independent spinor normalized as
\be\label{sumuu}
\sum_{\sigma}\bar{u} u(p,\sigma)=\slashed{p}+1 \;.
\ee
The positron Volkov solution is given by $\psi_\LCm(p',x)=\bar{K}(p',\phi)v(p',\sigma')\varphi(-p',x)$, where $\bar{K}(P,\phi)=K(-P,\phi)$ and 
\be\label{sumvv}
\sum_{\sigma'}\bar{v} v(p',\sigma')=\slashed{p}'-1 \;.
\ee
The propagator is given by
\be
S(x,y)=i\int\frac{\ud^4P}{(2\pi)^4}K\varphi(P,x)\frac{1}{\slashed{P}-1+i\epsilon}\bar{K}\varphi^*(P,y) \;.
\ee
We separate the propagator as~\cite{Seipt:2012tn}
\be
\frac{1}{\slashed{P}-1+i\epsilon}=\frac{1}{4P_\LCm}\left(\gamma^\LCp+\frac{\slashed{P}_{\rm on}+1}{P_\LCp-P^{\rm on}_\LCp+i\epsilon\,\text{sign}(P_\LCm)}\right) \;,
\ee
where $P^{\rm on}_\LCp=(1+P_\LCperp^2)/(4P_\LCm)$, and then the $P_\LCp$ integral gives
\be\label{PpInt}
\begin{split}
&\frac{i}{k_\LCp}\int\frac{\ud P_\LCp}{2\pi}\frac{e^{-i(x_2^\LCp-x_1^\LCp)P_\LCp}}{\slashed{P}-1+i\epsilon}=\frac{e^{-i(x_2^\LCp-x_1^\LCp)P^{\rm on}_{\LCp}}}{2kP}\Big\{i\slashed{k}\delta(\theta_{21}) \\
&+(\slashed{P}_{\rm on}+1)[\theta(kP)\theta(\theta_{21})-\theta(-kP)\theta(\theta_{12})]\Big\} \;,
\end{split}
\ee
where $\theta_{ij}=\phi_i-\phi_j$, $\phi_i=kx_i$.
The integrals over $x_{1,2}^{\LCm,\LCperp}$ give delta functions.
For $M_e$ we have $\bar{P}=\bar{p}_1:=\bar{p}+\bar{l}'$ ($kP>0$) and
for $M_p$ we have 
$\bar{P}=-\bar{p}_{\bar{1}}:=-(\bar{p}'+\bar{l}')$ ($kP<0$).

The total probability is given by
\be
\begin{split}
\mathbb{P}&=\frac{1}{2}\sum_{\rm spins}\int\!\ud\tilde{p}\ud\tilde{p}'\ud\tilde{l}'\left|\int\ud\tilde{l}f\frac{1}{k_\LCp}\bar{\delta}(p+p'+l'-l)M\right|^2 \\
&=\frac{1}{2}\sum_{\rm spins}\int\!\ud\tilde{p}\ud\tilde{p}'\frac{\theta(kl')}{klkl'}|M|^2 \;,
\end{split}
\ee
where $\bar{l}'=\bar{l}-\bar{p}-\bar{p}'$ and we have $1/2$ because we are averaging over the polarization of the initial photon. To compare with our results for trident and double Compton we introduce the following notation. We call the terms coming from $|M_e|^2$ and $|M_p|^2$ the direct part, and $\mathbb{P}_{\rm dir}=\mathbb{P}^e_{\rm dir}+\mathbb{P}^p_{\rm dir}$, where
\be
\mathbb{P}^{e,p}_{\rm dir}=\frac{1}{2}\sum_{\rm spins}\int\!\ud\tilde{p}\ud\tilde{p}'\frac{\theta(kl')}{klkl'}|M_{e,p}|^2 \;.
\ee
We refer to the cross-term as the exchange part
\be
\mathbb{P}_{\rm ex}=\frac{1}{2}\sum_{\rm spins}\int\!\ud\tilde{p}\ud\tilde{p}'\frac{\theta(kl')}{klkl'}2\text{Re}M_e^*M_p \;.
\ee

The integrals over $p_\LCperp$ and $p'_\LCperp$ are Gaussian and we perform them analytically for arbitrary field shape and polarization~\cite{Dinu:2013hsd}. We are left with the longitudinal momentum spectrum $\mathbb{P}(s)$, which we define as
\be
\mathbb{P}=:\int_0^{q_1}\ud s_0\ud s_2\theta(q_2)\mathbb{P}(s) \;,
\ee 
where we use the following notation for the longitudinal momenta, $s_0=kp/b_0$, $s_2=kp'/b_0$, $q_1=kl/b_0$ and $q_2=kl'/b_0=q_1-s_0-s_2$. When evaluating the spectrum we set $b_0=kl$, i.e. $q_1=1$. However, in order to see the symmetries and relation with double Compton we will keep $q_1$ explicit. For the momentum of the intermediate fermion we use $s_1=kp_1/b_0=q_1-s_2$ and $s_{\bar{1}}=kp_{\bar{1}}/b_0=q_1-s_0$.

\section{Exact results}\label{Exact results section}

Because of the separation of the propagator in~\eqref{PpInt}, the terms $\mathbb{P}^e_{\rm dir}$, $\mathbb{P}^p_{\rm dir}$ and $\mathbb{P}_{\rm ex}$ are each separated into three terms with two, three and four lightfront time integrals. 
To express these terms compactly we use the following definitions. For the longitudinal momenta we use $r_{ij}=(1/s_i)-(1/s_j)$, $\tilde{r}_{ij}=(1/s_i)+(1/s_j)$ and $\kappa_{ij}=(s_i/s_j)+(s_j/s_i)$. The field enters the exponential part of the integrands via the effective mass $M$~\cite{Kibble:1975vz},
\be
M^2_{ij}=1+\langle{\bf a}^2\rangle_{ij}-\langle{\bf a}\rangle_{ij}^2 \;,
\ee
where
\be
\langle F\rangle_{ij}=\frac{1}{\theta_{ij}}\int_{\phi_j}^{\phi_i}\!\ud\phi\; F(\phi) \;.
\ee
We also use $\Theta_{ij}=\theta_{ij}M^2_{ij}$.
The pre-exponential parts of the integrands can be expressed in terms of
\be\label{DeltaDefinition}
{\bf\Delta}_{ij}={\bf a}(\phi_i)-\langle{\bf a}\rangle_{ij} \;.
\ee
There are at most four $\phi$ integrals. We use $\phi_2$ and $\phi_4$ for the amplitude $M$ and $\phi_1$ and $\phi_3$ for its complex conjugate $M^*$. The Gaussian integrals over $p_\LCperp$ and $p'_\LCperp$ need to be regulated, which we do by replacing $\phi_{2,4}\to\phi_{2,4}+i\epsilon/2$ and $\phi_{1,3}\to\phi_{1,3}-i\epsilon/2$ where $\epsilon>0$. We leave the factors of $\epsilon$ implicit, as this can anyway be seen as a shift in the integration contours for $\phi_i$.  

For the direct terms we find
\be\label{Pe11exact}
\mathbb{P}^e_{11}(s)=\frac{\alpha^2}{4\pi^2}\frac{s_0s_2}{q_1^2s_1^2}\int\ud\phi_{12}\frac{-1}{\theta_{21}^2}e^{\frac{i}{2b_0}\tilde{r}_{20}\Theta_{21}} \;,
\ee 
where $\ud\phi_{12}=\ud\phi_1\ud\phi_2$,
\be\label{Pe12exact}
\begin{split}
\mathbb{P}^e_{12}(s)=\text{Re}\frac{i\alpha^2}{4\pi^2b_0q_1^2}\int&\frac{\ud\phi_{123}\theta(\theta_{31})}{s_1^3\theta_{21}\theta_{23}}e^{\frac{i}{2b_0}[\tilde{r}_{21}\Theta_{21}+r_{01}\Theta_{23}]} \\
&(q_1q_2-s_0s_2{\bf\Delta}_{12}\!\cdot\!{\bf\Delta}_{32}) \;,
\end{split}
\ee
and
\be\label{Pe22exact}
\begin{split}
\mathbb{P}^e_{22}(s)&=-\frac{\alpha^2}{4\pi^2b_0^2q_1^2}\int\frac{\ud\phi_{1234}\theta(\theta_{31})\theta(\theta_{42})}{s_1^2\theta_{21}\theta_{43}} \\
&\hspace{3cm}e^{\frac{i}{2b_0}[\tilde{r}_{21}\Theta_{21}+r_{01}\Theta_{43}]} \\
&\bigg\{\left[\frac{\kappa_{21}}{2}\left(\frac{2ib_0}{\tilde{r}_{21}\theta_{21}}+1+{\bf\Delta}_{12}\!\cdot\!{\bf\Delta}_{21}\right)+1\right] \\
&\times\left[\frac{\kappa_{01}}{2}\left(\frac{2ib_0}{r_{01}\theta_{43}}+1+{\bf\Delta}_{34}\!\cdot\!{\bf\Delta}_{43}\right)-1\right] \\
&-\frac{q_1q_2}{4s_1^2}\bigg[({\bf\Delta}_{21}-{\bf\Delta}_{12})\!\cdot\!({\bf\Delta}_{43}-{\bf\Delta}_{34})+ \\
&\frac{(s_0+s_1)(s_2-s_1)}{s_0s_2}({\bf\Delta}_{12}\!\times\!{\bf\Delta}_{21})\!\cdot\!({\bf\Delta}_{34}\!\times\!{\bf\Delta}_{43})\bigg]\bigg\} \;.
\end{split}
\ee
The corresponding terms for $\mathbb{P}^p$ can be obtained by replacing $s_0\leftrightarrow s_2$ (which means e.g. $s_1\leftrightarrow s_{\bar{1}}$).
As in the trident and the double Compton cases, we split the step functions as~\cite{Dinu:2017uoj}
\be
\begin{split}
\theta(\theta_{42})\theta(\theta_{31})=&\theta(\sigma_{43}-\sigma_{21})\Bigg\{1 \\
&-\theta\left(\frac{|\theta_{43}-\theta_{21}|}{2}-[\sigma_{43}-\sigma_{21}]\right)\Bigg\} \;,
\end{split}
\ee 
where the first term gives $\mathbb{P}^e_{22\to2}$, which we call the two-step part, and the second term gives $\mathbb{P}^e_{22\to1}$, which contributes to the one-step terms.

The two-step part can be obtained with the gluing approach presented in~\cite{Dinu:2019pau}. (In fact, the entire $\mathbb{P}_{22}^e$ can be obtained by including $\theta(\theta_{42})\theta(\theta_{31})$ instead of just $\theta(\sigma_{43}-\sigma_{21})$ in the integrand.) For photon trident we need all three $\mathcal{O}(\alpha)$ processes: nonlinear Breit-Wheeler and Compton scattering by either an electron or a positron. The spin and polarization structure of each of these can be expressed compactly in terms of the Stokes vectors, ${\bf n}_i$, for the initial and the two final-state particles, 
\be\label{PCnnn}
\begin{split}
\mathbb{P}=&\langle\mathbb{P}\rangle+{\bf n}_0\!\cdot\!{\bf P}_0+{\bf n}_1\!\cdot\!{\bf P}_1+{\bf n}_2\!\cdot\!{\bf P}_2 \\
&+{\bf n}_0\!\cdot\!{\bf P}_{01}\!\cdot\!{\bf n}_1+{\bf n}_0\!\cdot\!{\bf P}_{02}\!\cdot\!{\bf n}_2+{\bf n}_1\!\cdot\!{\bf P}_{12}\!\cdot\!{\bf n}_2 \\
&+{\bf P}_{012,ijk}{\bf n}_{0i}{\bf n}_{1j}{\bf n}_{2k}\;,
\end{split}
\ee
where the expressions for $\langle\mathbb{P}\rangle$ and ${\bf P}$ can be found in~\cite{Dinu:2019pau}. According to the gluing prescription, we have
\be
\mathbb{P}_{\rm glue}^e=2^4\langle\mathbb{P}_{\rm BW}\mathbb{P}_{\rm C}\rangle
\ee
and
\be
\mathbb{P}_{\rm glue}^p=2^4\langle\mathbb{P}_{\rm BW}\mathbb{P}_{\rm C}^{\rm p}\rangle \;,
\ee
where $\langle1\rangle=1$, $\langle{\bf n}\rangle=0$ and $\langle{\bf n}{\bf n}\rangle={\bf 1}$ for each particle, and there is a factor of $2^4$ because spin sums have been expressed in terms of averages for three final-state particles and one intermediate particle. In contrast to the trident case, see Eq.~(44) in~\cite{Dinu:2019pau}, there is no factor of $1/2$ since we do not have any identical particles here. By expressing the $\mathcal{O}(\alpha)$ processes as $\mathbb{P}={\bf N}_k^{(1)}{\bf N}_j^{(2)}{\bf M}_{kji}{\bf N}_i^{(0)}$, where ${\bf M}$ can be seen as a ``strong-field-QED Mueller matrix'', these averages $\langle...\rangle$ are equivalent to Mueller-matrix multiplication, ${\bf N}^{(3)}_m{\bf N}^{(2)}_l{\bf N}^{(1)}_j{\bf M}^{\rm C}_{mlk}{\bf M}^{\rm BW}_{kji}{\bf N}_i^{(0)}$.

For the exchange terms we find $\mathbb{P}_{\rm ex}^{11}=0$,
\be
\begin{split}
\mathbb{P}_{\rm ex}^{12}(s)=\text{Re}\frac{-i\alpha^2}{4\pi^2b_0q_1^2}\int&\frac{\ud\phi_{123}\theta(\theta_{31})}{s_{\bar{1}}\theta_{21}\theta_{23}}e^{\frac{i}{2b_0}[\tilde{r}_{21}\Theta_{21}+r_{01}\Theta_{23}]} \\
&{\bf\Delta}_{12}\cdot{\bf\Delta}_{32} \;,
\end{split}
\ee 
$\mathbb{P}_{\rm ex}^{21}(s)=\mathbb{P}_{\rm ex}^{12}(s)\big|_{s_0\leftrightarrow s_2}$,
and finally the most difficult term
\be
\begin{split}
\mathbb{P}_{\rm ex}^{22}(s)&=\text{Re}\frac{-\alpha^2}{8\pi^2b_0^2q_1^2}\int\frac{\ud\phi_{1234}\theta(\theta_{31})\theta(\theta_{42})}{s_0s_1s_{\bar{1}}s_2d_0} \\
&\exp\bigg\{\frac{iq_1q_2}{2b_0s_0s_1s_{\bar{1}}s_2d_0}\bigg[\theta_{21}\theta_{43}\left(\frac{\Theta_{21}}{q_2}-\frac{\Theta_{43}}{q_1}\right)+ \\
&\theta_{23}\theta_{41}\left(\frac{\Theta_{41}}{s_2}+\frac{\Theta_{23}}{s_0}\right)+\theta_{31}\theta_{42}\left(\frac{\Theta_{42}}{s_{\bar{1}}}-\frac{\Theta_{31}}{s_1}\right)\bigg]\bigg\} \\
&\bigg\{F_0+f_0+\frac{2ib_0}{d_0}(f_1+z_1)+\left(\frac{2b_0}{d_0}\right)^2z_2\bigg\} \;,
\end{split}
\ee
where
\be
d_0=\frac{\theta_{42}\theta_{31}}{s_1s_{\bar{1}}}+\frac{\theta_{23}\theta_{41}}{s_0s_2} \;,
\ee
\be
\begin{split}
F_0=&(\kappa_{02}+\kappa_{1\bar{1}})({\bf d}_1\!\cdot\!{\bf d}_2)({\bf d}_4\!\cdot\!{\bf d}_3) \\
&+(\kappa_{02}-\kappa_{1\bar{1}})({\bf d}_1\!\times\!{\bf d}_2)\!\cdot\!({\bf d}_4\!\times\!{\bf d}_3) \;,
\end{split}
\ee
\be
\begin{split}
f_0=&\frac{1}{s_0s_1s_{\bar{1}}s_2}\Big[(s_1q_2{\bf d}_1-s_{\bar{1}}q_1{\bf d}_4)\!\cdot\!(s_{\bar{1}}q_2{\bf d}_2-s_1q_1{\bf d}_3) \\
&+(s_2q_1{\bf d}_4+s_0q_2{\bf d}_2)\!\cdot\!(s_2q_2{\bf d}_1+s_0q_1{\bf d}_3)\Big] \;,
\end{split}
\ee
\be
\begin{split}
f_1=&\kappa_{02}\left[\frac{\theta_{41}}{s_2}{\bf d}_1\!\cdot\!{\bf d}_4+\frac{\theta_{23}}{s_0}{\bf d}_3\!\cdot\!{\bf d}_2\right]+ \\
&\kappa_{1\bar{1}}\left[-\frac{\theta_{31}}{s_1}{\bf d}_1\!\cdot\!{\bf d}_3+\frac{\theta_{42}}{s_{\bar{1}}}{\bf d}_2\!\cdot\!{\bf d}_4\right]+\\
&(\kappa_{02}+\kappa_{1\bar{1}})\left[-\frac{\theta_{21}}{q_2}{\bf d}_1\!\cdot\!{\bf d}_2+\frac{\theta_{43}}{q_1}{\bf d}_4\!\cdot\!{\bf d}_3\right] \;,
\end{split}
\ee
\be
\begin{split}
z_1=&\frac{q_1^2}{s_2s_1q_2}\left(3-\frac{s_{\bar{1}}s_0}{s_1s_2}\right)\phi_1-\frac{q_2^2}{s_2s_{\bar{1}}q_1}\left(3-\frac{s_1s_0}{s_{\bar{1}}s_2}\right)\phi_4+ \\
&
\frac{q_2^2}{s_1s_0q_1}\left(3-\frac{s_{\bar{1}}s_2}{s_1s_0}\right)\phi_3-\frac{q_1^2}{s_{\bar{1}}s_0q_2}\left(3-\frac{s_1s_2}{s_{\bar{1}}s_0}\right)\phi_2 \;,
\end{split}
\ee
\be
z_2=-\kappa_{02}\frac{\theta_{23}\theta_{41}}{s_0s_2}+\kappa_{1\bar{1}}\frac{\theta_{31}\theta_{42}}{s_1s_{\bar{1}}}+(\kappa_{02}+\kappa_{1\bar{1}})\frac{\theta_{43}\theta_{21}}{q_1q_2} \;,
\ee
and
\be
{\bf d}_1=\frac{q_2}{s_0s_{\bar{1}}d_0}\left[\frac{\theta_{21}\theta_{43}}{q_2}{\bf\Delta}_{12}-\frac{\theta_{31}\theta_{42}}{s_1}{\bf\Delta}_{13}+\frac{\theta_{23}\theta_{41}}{s_2}{\bf\Delta}_{14}\right]
\ee 
\be
{\bf d}_2=\frac{q_2}{s_1s_2d_0}\left[\frac{\theta_{21}\theta_{43}}{q_2}{\bf\Delta}_{21}+\frac{\theta_{23}\theta_{41}}{s_0}{\bf\Delta}_{23}-\frac{\theta_{31}\theta_{42}}{s_{\bar{1}}}{\bf\Delta}_{24}\right]
\ee 
\be
{\bf d}_3=\frac{q_1}{s_2s_{\bar{1}}d_0}\left[\frac{\theta_{42}\theta_{31}}{s_1}{\bf\Delta}_{31}+\frac{\theta_{23}\theta_{41}}{s_0}{\bf\Delta}_{32}-\frac{\theta_{43}\theta_{21}}{q_1}{\bf\Delta}_{34}\right]
\ee 
\be
{\bf d}_4=\frac{q_1}{s_0s_1d_0}\left[\frac{\theta_{23}\theta_{41}}{s_2}{\bf\Delta}_{41}+\frac{\theta_{42}\theta_{31}}{s_{\bar{1}}}{\bf\Delta}_{42}-\frac{\theta_{43}\theta_{21}}{q_1}{\bf\Delta}_{43}\right]
\ee

\subsection{Comparison with double Compton}

These expressions look very similar to the ones we derived in~\cite{Dinu:2018efz} for double Compton. In fact, we can obtain each of these terms by the following replacements.
 
Let us start with $\mathbb{P}^e$. To obtain $\mathbb{P}^e_{11}$, $\mathbb{P}^e_{12}$ and $\mathbb{P}^e_{22}$ in~\eqref{Pe11exact}, \eqref{Pe12exact} and~\eqref{Pe22exact} from the expression in~\cite{Dinu:2018efz} for the direct part of the probability of double Compton, note first that those expressions are expressed as $\mathbb{P}_{\rm dir}^{\rm DC}(q)=\mathbb{P}_{\rm dir,asym}^{\rm DC}(q_1,q_2)+\mathbb{P}_{\rm dir,asym}^{\rm DC}(q_2,q_1)$. To go from $\mathbb{P}_{\rm dir,asym}^{\rm DC}(q_1,q_2)$ to $\mathbb{P}^e$,      
replace $q_1\to-q_1$ (we change an outgoing photon to an incoming one), $s_0\to-s_2$ (the initial electron becomes an outgoing positron), and $s_2\to s_0$ (just different notation for the outgoing electron). 
These changes take care of all the nontrivial parts of the expressions. To obtain the correct overall factor we have to multiply by an overall factor of $-2/q_1^2$. The reason for this sign is that when changing an electron to a positron by replacing $p\to-p'$, the spin sum in~\eqref{sumuu} gives minus~\eqref{sumvv}. The factor of 2 is due to the fact that in double Compton one has to divide by 2 to prevent double counting of identical particles, while here there are no identical particles in the final state. The factor of $q_1^2$ is just normalization, and we anyway put $q_1=1$ when evaluating these expressions. The expressions for $\mathbb{P}^p$ can of course also be obtained in this way, since $\mathbb{P}^p$ can be obtained from $\mathbb{P}^e$ by replacing $s_0\leftrightarrow s_2$.  

To obtain $\mathbb{P}_{\rm ex}$ from the exchange terms, we first note that $\mathbb{P}_{\rm ex}^{12\rm DC}(q)=\mathbb{P}_{\rm ex,asym}^{12\rm DC}(q_1,q_2)+\mathbb{P}_{\rm ex,asym}^{12\rm DC}(q_2,q_1)$. $\mathbb{P}_{\rm ex}^{12}(s)$ is obtained from $\mathbb{P}_{\rm ex,asym}^{12\rm DC}(q_1,q_2)$ by replacing $q_1\to-q_1$, $s_0\to-s_2$ and $s_2\to s_0$, as for the direct terms, plus $s_{\bar{1}}\to-s_{\bar{1}}$ (the intermediate electron is changed to a positron). We also have to multiply by a factor of $-2/q_1^2$ for the same reasons as for $\mathbb{P}^e$.


For $\mathbb{P}_{\rm ex}^{22}$, we note that in the expressions above and in~\cite{Dinu:2018efz} we have named the $\phi$ variables such that the $\phi_2$ step happens before the $\phi_4$ step, which is why we have $\theta(\theta_{42})$. However, looking at the second line in~\eqref{PpInt}, we see that to compare with the double-Compton expressions in~\cite{Dinu:2018efz} it is more natural to rename the integration variables as $\phi_2\leftrightarrow\phi_4$ in the above expressions for $\mathbb{P}_{\rm ex}^{22}$. Then $\phi_1$ and $\phi_4$ are the vertices connected to the $q_1$-photon line and $\phi_2$ and $\phi_3$ are the vertices connected to the $q_2$-photon line, which is also how the photon lines are connected for the choice of variable names we made in the double-Compton case~\cite{Dinu:2018efz}. With this renaming, the $\phi_4$ vertex happens before the $\phi_2$ vertex, and $\mathbb{P}_{\rm ex}^{22}$ is obtained from $\mathbb{P}_{\rm ex}^{22\rm DC}$ with the same replacements for the momentum variables and overall prefactor as for the other terms, plus changing one of the step function $\theta(\theta_{42})\to-\theta(\theta_{24})$. This replacement of the step function and the extra sign change is due to the relative sign between the two terms in the second line in~\eqref{PpInt}, which in turn comes from the fact that the $P_\LCp$ integration contour should be closed in the upper and lower complex plane for $P_\LCm>0$ and $P_\LCm<0$, respectively. So, $\theta(\theta_{42})\to-\theta(\theta_{24})$ is a consequence of the replacements of the momentum variables.
We could trivially make the same $\phi_2,\phi_4$ replacement for the entire exchange part, because this does not have any effect on $\mathbb{P}_{\rm ex}^{12}$, which can be seen as a 4D integral with $\delta(\theta_{42})$ instead of $\theta(\theta_{42})$ or $\theta(\theta_{24})$.   

\begin{figure*}
\includegraphics[width=\linewidth]{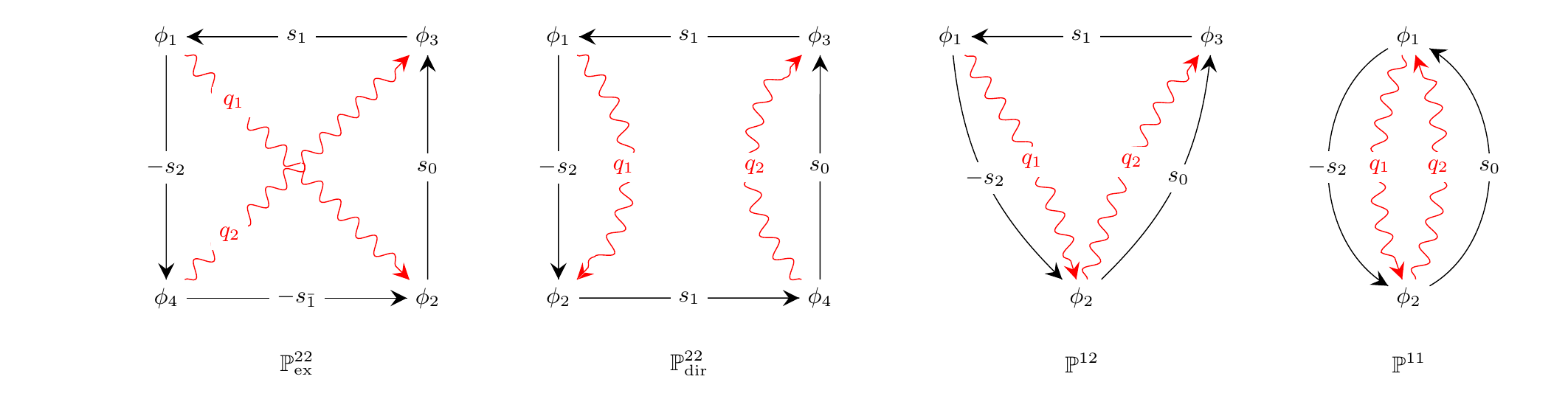}
\caption{Probability diagrams~\cite{Dinu:2017uoj,Dinu:2018efz} for photon trident.}
\label{probDiagramFig}
\end{figure*}

These relations can be better understood by comparing the probability diagrams in Fig.~\ref{probDiagramFig} with the corresponding ones in~\cite{Dinu:2018efz}. 

\subsection{Saddle-point approximations}\label{Saddle point approximations section}         

In this section we will derive saddle-point approximations for the emission of a hard photon and $\chi\ll1$. We consider linearly polarized fields, $a(\phi)=a_0f(\phi)$. In the LCF regime we have $a_0\gg1$ and we can expand the probability in a series in $1/a_0$. We perform the $\phi$ integrals with the saddle-point method. The calculations are almost identical to the ones in~\cite{Dinu:2018efz}, so we simply state the results. We find 
\be
\mathbb{P}_{11}^e=\frac{\alpha^2}{8\pi^\frac{3}{2}q_1^2}\frac{s_0s_2}{s_1^2\sqrt{\tilde{r}_{20}}}\int\frac{\ud\phi}{b_0}\chi^\frac{3}{2}e^{-\frac{2\tilde{r}_{20}}{3\chi}} \;,
\ee
where $\chi(\phi)=a_0b_0f'(\phi)$,
\be
\mathbb{P}_{12}^e=\frac{\alpha^2}{24\pi^\frac{3}{2}q_1^2}\frac{4q_1q_2+s_0s_2}{s_1^3\sqrt{\tilde{r}_{20}}}\left[\frac{1}{\tilde{r}_{21}}-\frac{1}{r_{01}}\right]\int\frac{\ud\phi}{b_0}\chi^\frac{3}{2}e^{-\frac{2\tilde{r}_{20}}{3\chi}} \;,
\ee
\be
\mathbb{P}_{\rm ex}^{12}=\frac{\alpha^2}{24\pi^\frac{3}{2}q_1^2}\frac{1}{s_{\bar{1}}\sqrt{\tilde{r}_{20}}}\left[\frac{1}{\tilde{r}_{21}}-\frac{1}{r_{01}}\right]\int\frac{\ud\phi}{b_0}\chi^\frac{3}{2}e^{-\frac{2\tilde{r}_{20}}{3\chi}} \;,
\ee
\be\label{P221saddleLCF}
\mathbb{P}_{22\to1}^e=\frac{-\alpha^2}{4\pi^\frac{3}{2}q_1^2}\sqrt{\tilde{r}_{20}}\left[\frac{q_1}{q_2}+\frac{q_2}{q_1}-\frac{s_1s_{\bar{1}}}{q_1q_2}\right]\int\frac{\ud\phi}{b_0}\sqrt{\chi}e^{-\frac{2\tilde{r}_{20}}{3\chi}} \;,
\ee
\be
\begin{split}
\mathbb{P}_{22\to2}^e=&\frac{\alpha^2}{4\pi q_1^2}\sqrt{\frac{q_1q_2}{s_0s_2}}\frac{1}{s_1}\left[\frac{q_1}{q_2}+\frac{q_2}{q_1}-\frac{s_1s_{\bar{1}}}{q_1q_2}\right] \\
&\int\frac{\ud\sigma_1}{b_0}\int_{\sigma_1}\frac{\ud\sigma_2}{b_0}\sqrt{\chi(\sigma_1)\chi(\sigma_2)}e^{-\frac{2\tilde{r}_{21}}{3\chi(\sigma_1)}-\frac{2r_{01}}{3\chi(\sigma_2)}} 
\end{split}
\ee
and
\be\label{exdircancelsaddle}
\mathbb{P}_{\rm ex}^{22}=-\mathbb{P}_{22\to1} \;,
\ee   
where $\mathbb{P}_{22\to1}:=\mathbb{P}_{22\to1}^e+\mathbb{P}_{22\to1}^p$ (note that~\eqref{P221saddleLCF} is symmetric in $s_0\leftrightarrow s_2$, so $\mathbb{P}_{22\to1}^e=\mathbb{P}_{22\to1}^p$ to leading order). 
In fact, these expressions can be obtained from the corresponding results in~\cite{Dinu:2018efz} by simply making the replacements as explained in the previous section.
Thus, the exchange term cancels the direct part of the one-step to leading order, not just in the double Compton case, but also for photon trident. This tells us that the two-step is a better approximation of the total probability than what the scaling $\mathbb{P}_{\rm two}\sim\mathcal{O}(a_0^2)$, $\mathbb{P}_{\rm one}\sim\mathcal{O}(a_0)$ alone suggests.

The above results holds for arbitrary field shapes with $a_0\gg1$. There are certain field shapes which also allow us to obtain simple expressions for $a_0\sim1$. One such example is a Sauter pulse, $a(\phi)=a_0\tanh\phi$. The results for photon trident agree with what one obtains by making the replacements of the longitudinal momenta in the corresponding results in~\cite{Dinu:2018efz}. 
Thus, for these single-maximum fields, all three second-order processes with one initial particle (trident, double Compton, and photon trident) have the same $a_0$ dependence in the leading, exponential part of the probability, only the dependencies on the longitudinal momenta are different.    

If we use $q_2$ and $s_0$ as independent integration variables, then we can perform the $s_0$ integral with the saddle-point method. The saddle point is given by $s_0=s_2=(q_1-q_2)/2$ and, for $a_0\gg1$, we find that the probability scales as 
\be
\mathbb{P}\sim\exp\left\{-\frac{8}{3(q_1-q_2)\chi}\right\} \;,
\ee
so emitting a hard photon ($q_2\sim q_1$) leads to an increased exponential suppression compared to the Breit-Wheeler case, as expected.

\section{Resummation of small-$\chi$ expansion}\label{resumSmallSection}

In the previous section we used the leading-order in the saddle-point expansion to see how important various terms are. We saw in particular that the exchange term cancels the direct part of the one-step to leading order. This means that one has to go beyond the leading order for these terms. However, the small-$\chi$ expansion is asymptotic and, as we showed in~\cite{Dinu:2018efz} for double Compton, for these processes the region where the precision is improved by adding the first couple of next-to-leading order terms is limited to so small $\chi$ that the exponential suppression makes results very small. Fortunately, as demonstrated in~\cite{Torgrimsson:2020wlz} for trident, one can use resummation methods to resum these asymptotic series.  

The resummations methods that we will describe in the following sections are quite general.
We focus initially on double Compton scattering as an example, and return to photon trident in Sec.~\ref{photonTridentResum}. In~\cite{Dinu:2018efz} we plotted the probability as a function of $\chi$ for several different values of $q_1$ and $q_2$. We have checked that the resummations presented below agree with the numerical results from~\cite{Dinu:2018efz}. However, we will present several different resummations of both the small- and the large-$\chi$ expansions, which means that we do not need exact numerical results in order to check the precision of the resummations. Instead, to check, for example, the precision of the small-$\chi$ resummations at large $\chi$ we can use the large-$\chi$ expansion, and vice versa for large-$\chi$ resummations. There is also a large interval around $\chi\sim1$ where the small- and large-$\chi$ expansions agree to a high precision. So, in the plots below, the exact numerical result is not included, but if it were it would in all cases be indistinguishable from at least one of the curves. 

From~\cite{Dinu:2018efz} we see that in some cases there is a large degree of cancellation between the direct and exchange parts of the one-step, not just for small $\chi$, but also as $\chi$ becomes large. One typical example where this happens is $q_1=q_2=1/3$, which we use as a first example. Here the one-step terms can be expanded as
\be\label{dirOneSmallExpansion}
\mathbb{P}_{\rm dir}^{\rm one}=-\frac{3\alpha^2a_0\Delta\phi}{\sqrt{2}\pi^{3/2}\sqrt{\chi}}\exp\left(-\frac{4}{3\chi}\right)T^{\rm dir} \;,
\ee
\be\label{exOneSmallExpansion}
\mathbb{P}_{\rm ex}^{\rm one}=\frac{3\alpha^2a_0\Delta\phi}{\sqrt{2}\pi^{3/2}\sqrt{\chi}}\exp\left(-\frac{4}{3\chi}\right)T^{\rm ex} \;,
\ee
where 
\be
T=\sum_{n=0}^\infty T_n\chi^n 
\ee
\be
T^{\rm dir}=1+\frac{1907}{864}\chi-\frac{18761023}{1492992}\chi^2+\frac{51512914979}{429981696}\chi^3-... 
\ee 
\be
\begin{split}
T^{\rm ex}=&1+\frac{50021}{30240}\chi-\frac{48618935483}{7472424960}\chi^2 \\
&+\frac{231504152856583}{4860943073280}\chi^3-... \;.
\end{split}
\ee
A direct sum of the series in $T$ does not work, because this is an asymptotic series with factorially growing coefficients. The standard approach for such series is to use Borel transformation,
\be
BT(t)=\sum_{n=0}^\infty B_n t^n=\sum_{n=0}^\infty\frac{T_n}{n!}t^n \;.
\ee 
By calculating a finite number of terms we obtain a truncated Borel transform $BT_N=\sum_{n=0}^NB_nt^n$, which needs to be resummed before we transform back to $\chi$. This can be done with Borel-Pad\'e-conformal methods~\cite{Guillou1980,Caliceti:2007ra,Costin:2019xql,Costin:2020hwg,KleinertPhi4,ZinnJustinBook,Florio:2019hzn,Baker1961,BenderOrszag}.

\begin{figure}
\includegraphics[width=\linewidth]{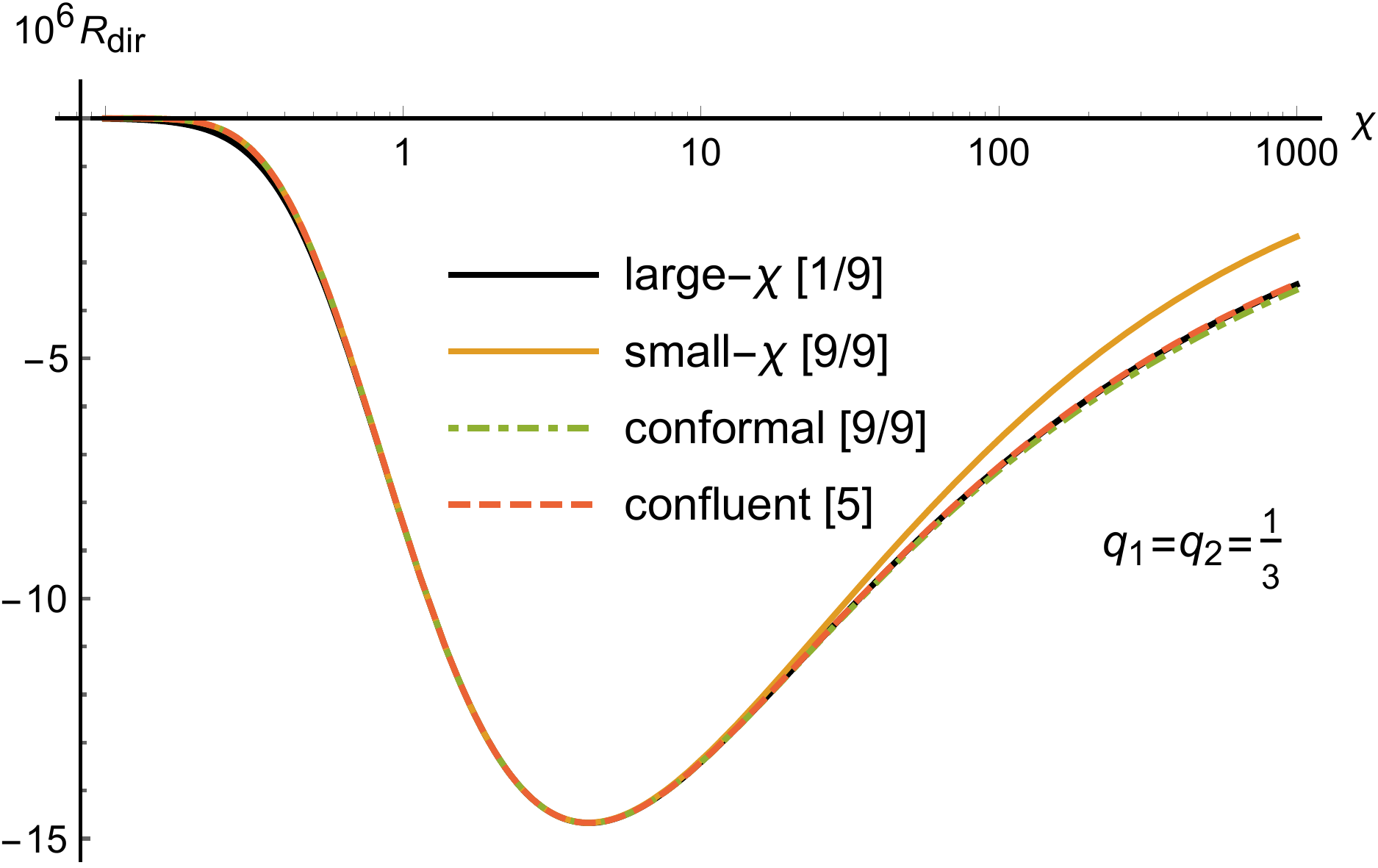}
\includegraphics[width=\linewidth]{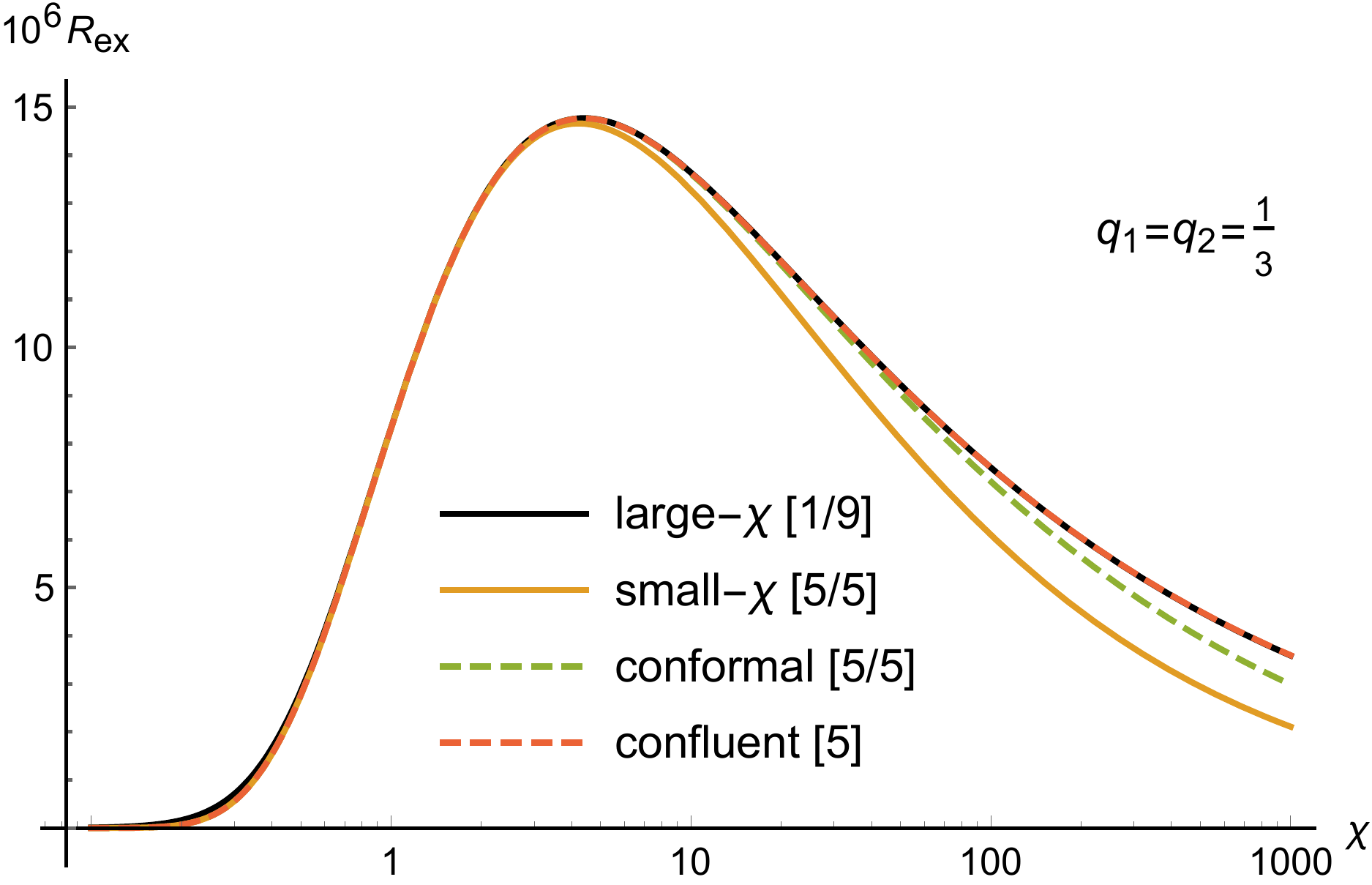}
\caption{Comparison of different resummation methods for the direct and exchange parts of the one-step, $R_{\rm dir}=\mathbb{P}_{\rm dir}^{\rm one}/(a_0\Delta\phi)$ and $R_{\rm ex}=\mathbb{P}_{\rm ex}^{\rm one}/(a_0\Delta\phi)$. For the confluent resummation we have chosen $b=19/20$ ($b=2/5$) for the direct (exchange) part. However, in this case this resummation is not very sensitive to the value of $b$, and the difference from e.g. $b=1$ is on the order of the width of these lines at $\chi=1000$.} 
\label{DCdir1d3resumFig}
\end{figure}

One way is to calculate a Pad\'e approximant~\cite{Baker1961,BenderOrszag,KleinertPhi4,ZinnJustinBook} for the truncated series
\be
PBT[m/n](t)=\frac{\sum_{i=0}^m A_i t^i}{1+\sum_{j=1}^n B_j t^j}=BT_N(t)+\mathcal{O}(t^{N+1}) \;,
\ee 
where $m$ and $n$ are integers with $m+n+1\leq N$. The Pad\'e approximant provides an analytic continuation of the truncated series beyond its (finite) radius of convergence. The final step is to take the inverse of the Borel transform, i.e. the Laplace transform, 
\be
PBT[m/n](\chi)=\int_0^\infty\frac{\ud t}{\chi}e^{-t/\chi}PBT[m/n](t) \;,
\ee  
which gives a resummation of the original $T$ series.

In practice, it can for some contributions be challenging to calculate a large number of terms. For these sort of asymptotic series obtained with the saddle-point method, the challenge is that the number of different terms in the integrand can become very large if there are several integration variables. 
So, one needs to make the most of the terms one has. One common way to improve the convergence is to make a conformal transformation before making the Pad\'e approximant~\cite{Guillou1980,Costin:2019xql,Costin:2020hwg,Caliceti:2007ra,KleinertPhi4,ZinnJustinBook}. For this one uses additional information about the series, in particular, the position of the singularity closest to the origin. By calculating the first $\sim20$ terms (for the direct part, $\sim10$ for the exchange part) and by matching the ratios of neighboring coefficients of the Borel transform onto $B_{n+1}/B_n=c_0+c_1/n+c_2/n^2+...$, we find (for this example) $c_0=-3$ and hence the Borel transform has a finite radius of convergence limited by a singularity at $t=t_0=-1/3$. We make a change of variable in $BT_N(t)$ from $t$ to the conformal variable $z$~\cite{optimalConformal},
\be
z=\frac{\sqrt{1+\frac{t}{t_0}}-1}{\sqrt{1+\frac{t}{t_0}}+1} \quad t=\frac{4t_0z}{(1-z)^2} \;,
\ee  
which maps $t<t_0$ onto the unit circle in the complex $z$ plane. The resulting function is then re-expanded in powers of $z$ to the same order as $BT_N(t)$. Next, one makes a Pad\'e approximant $[m/n](z)$ of this series in $z$. Expressing $z$ in terms of $t$ gives a Pad\'e-conformally resummed Borel transform $PCBT(t)$, and then the last step is to perform the Laplace transform,
\be
PCBT[m/n](\chi)=\int_0^\infty\frac{\ud t}{\chi}e^{-t/\chi}PCBT[m/n](t) \;.
\ee

Another resummation method was proposed in~\cite{Alvarez:2017sza}, which we found to be very useful for resumming saddle-point series for the two-step part of trident and nonlinear Breit-Wheeler in~\cite{Torgrimsson:2020wlz}. It is even more useful here, because we have access to fewer terms in the small-$\chi$ expansion.
In this method one makes use of the additional information about the scaling at large $\chi$. The transform is given by a linear superpostion of a certain function $\phi(x)$ with rescaled argument,
\be\label{ASresumDefinition}
AST_n(\chi)=\sum_{i=1}^n\frac{c_i}{-\chi_i}\phi\left(-\frac{\chi}{\chi_i}\right) \;.
\ee 
The constants $c_i$ and $\chi_i$ are obtained by demanding that the first $2n$ terms in the series in $\chi$ match the terms in the series to be resummed. Although one can choose different functions, we will choose the confluent hypergeometric function suggested in~\cite{Alvarez:2017sza},
\be\label{Uab}
\phi(x)=x^{-a}U\left(a,1+a-b,\frac{1}{x}\right) \;,
\ee 
where $a$ and $b$ are two constants. We will show below that $\mathbb{P}_{\rm dir}^{\rm one}$ scales as $1/\chi^{1/3}$ for large $\chi$, which means that we want the resummed $T$ to scale as $\chi^{1/6}$. To match this large-$\chi$ scaling we choose $a=-1/6$. The second constant, $b$, is not determined by this scaling. In some cases one can obtain a significant improvement at large $\chi$ by choosing a suitable value of $b$, which one can find by testing a couple of different values and see which leads to the best agreement with the large-$\chi$ expansion at large $\chi$ (which we will derive in the next section)\footnote{One could of course imagine a more precise, numerical determination of $b$ by comparing with the large-$\chi$ result. However, for the examples we have considered, it is usually enough to try only a couple of different values of $b$ in order to obtain a resummation that is indistinguishable from the exact result on the scale of the plot (assuming, of course, that enough terms have been calculated and that such a value of $b$ exists). Moreover, below we will anyway present new resummation methods that allow us to resum the small-$\chi$ expansion using several (in principle arbitrarily many) terms from the large-$\chi$ expansion.}.  
The fact that the large-$\chi$ scaling is built into the resummation allows us to obtain a good precision (even at large $\chi$) with fewer terms than what is needed for the conformal-Pad\'e method.

In Fig.~\ref{DCdir1d3resumFig} we compare these resummation methods. Consider first the direct part, for which it is easier to obtain more terms. We see that the Pad\'e resummation with $[9/9]$ gives a good precision up to $\chi\sim20$, while for larger $\chi$ there is a small difference from the exact result. With the Pad\'e-conformal method with $[9/9]$ we find good precision for the entire range plotted. An even better precision can be obtained with only $\sim$ half the number of terms using the confluent resummation $AST_5$.   

For the exchange part, $\mathbb{P}_{\rm ex}^{22}$, it can be challenging/time-consuming to obtain a large number of terms. Here we have calculated the first 11 terms, i.e. up to $\chi^{10}$. For the Pad\'e and Pad\'e-conformal methods, this allows us to use $[5/5]$. As we see in Fig.~\ref{DCdir1d3resumFig}, this means that the Pad\'e resummation breaks down sooner, at $\chi\gtrsim10$, and the relative error is larger in the large-$\chi$ part of the plotted interval. The Pad\'e-conformal resummation is still rather good even for larger $\chi$, with a relative error of $\sim4\%$ at $\chi\sim100$. We can still use the same order for the confluent resummation, i.e. $AST_5$, which still gives a very precise result even at very large $\chi$. 

It is natural from a calculational point of view to consider the direct and exchange parts separately. For example, here we can calculate many more terms for the direct part, which is good for both precision as well as in order to determine the singularities of the Borel transform. However, the direct and exchange parts are in general on the same order of magnitude and (for double Compton scattering, but not for trident) only their sum is gauge invariant. For some values of $q_1$ and $q_2$, the direct and exchange parts only cancel each other (to leading order) for small $\chi$, but not for large $\chi$. For such cases we could simply resum the direct and exchange parts separately before adding them together, which would give a relative error for their sum on the same order of magnitude as the relative error for the direct and exchange parts separately. However, here we have chosen a more challenging example, where $|\mathbb{P}_{\rm dir}^{\rm one}+\mathbb{P}_{\rm ex}^{\rm one}|$ is much smaller than $|\mathbb{P}_{\rm dir}^{\rm one}|$ and $|\mathbb{P}_{\rm ex}^{\rm one}|$, which means that even a small error in the resummation of the terms separately can be a large error for their sum. In such cases it is better to first add together the two $\chi$ series before resumming. The resummation of $\mathbb{P}_{\rm dir}^{\rm one}+\mathbb{P}_{\rm ex}^{\rm one}$ with the confluent hypergeometric method is shown in Fig.~\ref{largechiexpFig}, where we find a very good agreement over essentially the entire $\chi$ range by a suitable choice of $a$ and $b$ in~\eqref{Uab}. However, to make this choice we first have to derive the large-$\chi$ expansion.

\section{Resummation of large-$\chi$ expansion}\label{resumLargeSection}

\begin{figure}
\includegraphics[width=\linewidth]{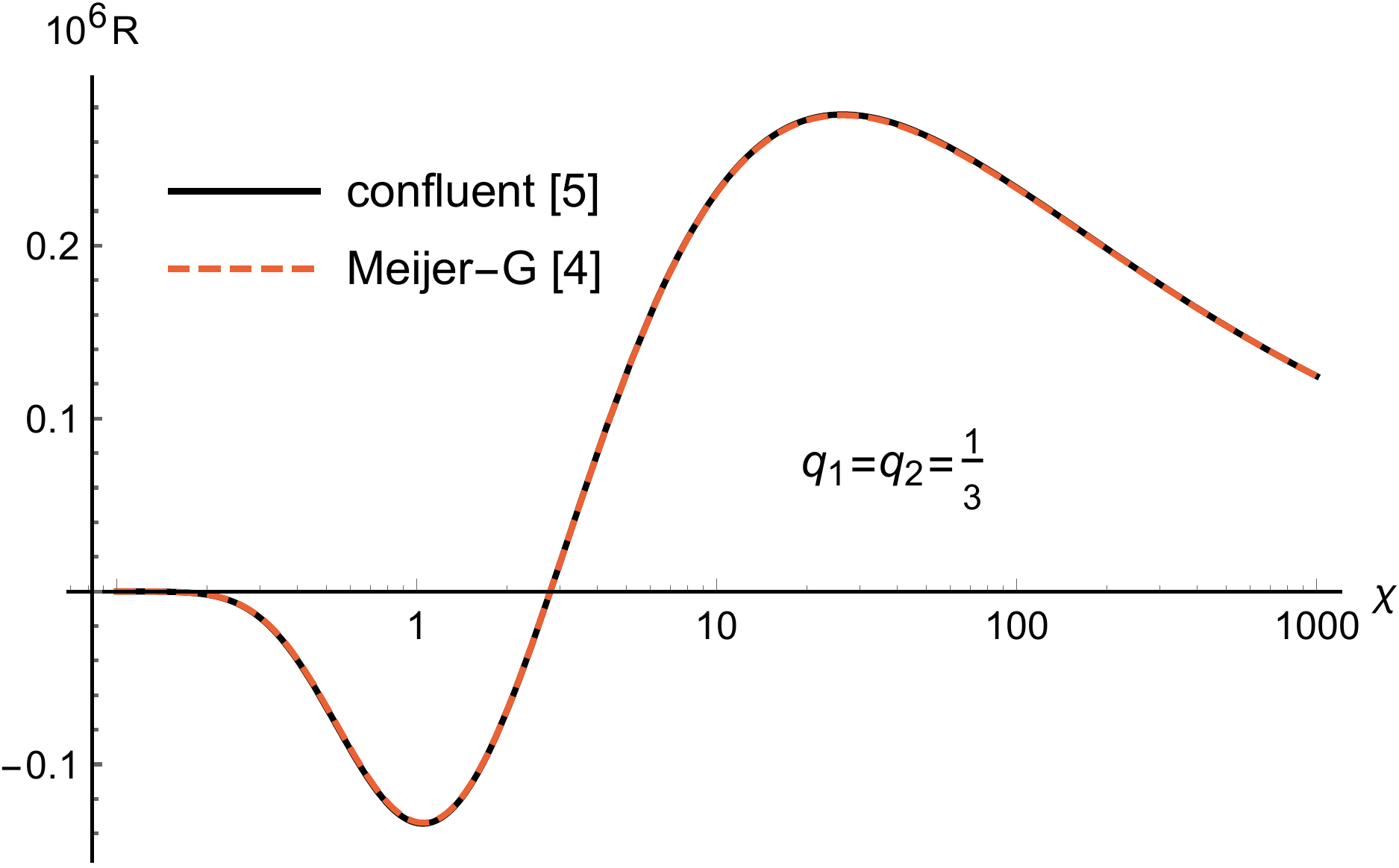}
\includegraphics[width=\linewidth]{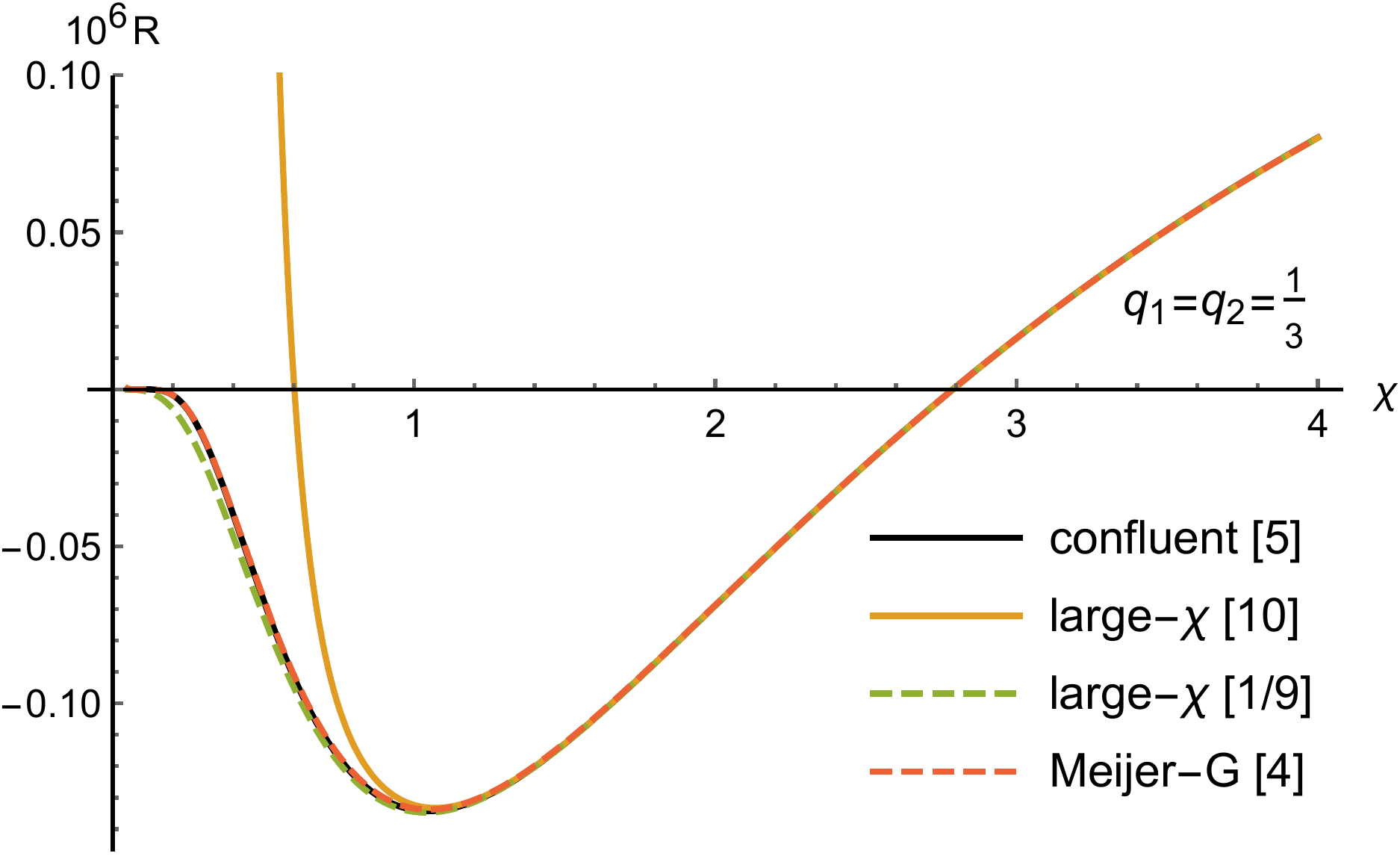}
\caption{Comparison of the different large-$\chi$ resummations for the total one-step. On this scale and in this interval, the exact result is well approximated by the confluent hypergeometric resummation in~\eqref{ASresumDefinition} with $n=5$, $a=5/6$ and $b=156/100$, where $a$ follows from the leading large-$\chi$ scaling ($1/\chi^{2/3}$) and the value of $b$ is the result of a very rough optimization at large $\chi$, where the result is very well approximated by the large-$\chi$ expansion.
The ``large-$\chi$ $[10]$'' line is the result of a direct summation of the first 11 terms in~\eqref{PoneLargeExpansion}; the ``large-$\chi$ $[1/9]$'' line is a $[1/9]$ Pad\'e approximant of that sum; and ``Meijer-G'' is the new resummation in~\eqref{sumofG} with $c_n$, $n=0,...,4$, determined by matching its large-$\chi$ expansion with the first 5 terms in~\eqref{PoneLargeExpansion}.}
\label{largechiexpFig}
\end{figure}

\begin{figure}
\includegraphics[width=\linewidth]{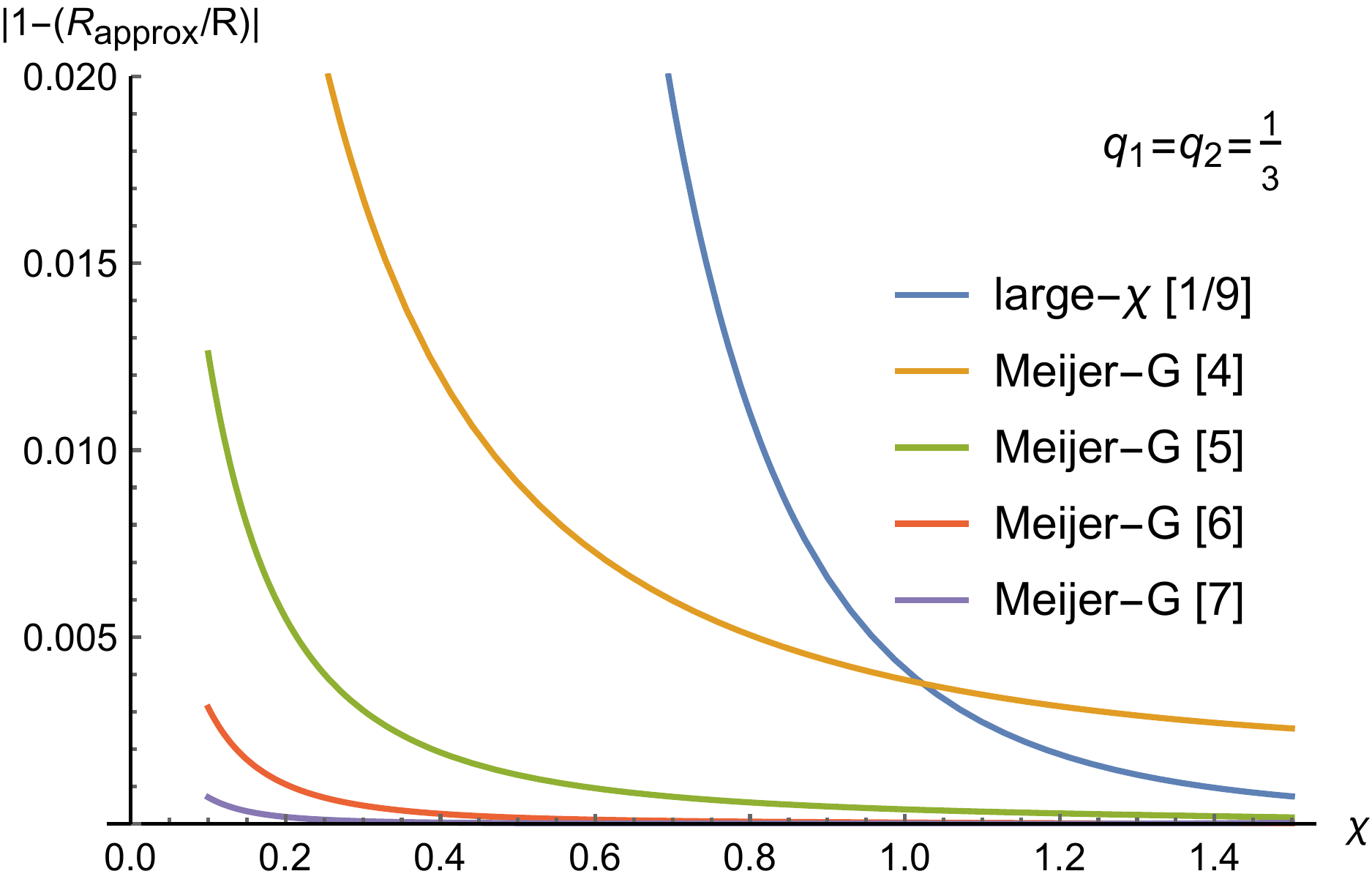}
\caption{Relative error of large-$\chi$ resummations. Same notation as in Fig.~\ref{largechiexpFig}. At $\chi=1$ the probability is exponentially small $\mathbb{P}_{\rm one}\sim10^{-12}$ compared to Fig.~\ref{largechiexpFig}. To estimate the relative error, the exact result has been approximated by the confluent hypergeometric resummation in Fig.~\ref{largechiexpFig}, which is possible because its relative error is much smaller in this $\chi$ interval (it is after all a resummation of the small-$\chi$ expansion and hence becomes more precise as $\chi$ decreases).}
\label{MGresumRelErFig}
\end{figure}

In this section we will consider the large-$\chi$ expansion. To obtain this expansion we first need to know how the integration variables scale. In the simplest term, $\mathbb{P}_{11}$, the exponential part of the integrand is given by
\be
\exp\left\{\frac{ir\theta}{2\chi}\left(1+\frac{\theta^2}{12}\right)\right\} \;,
\ee
which has been obtained from the integrand for the exact result by rescaling $\theta\to\theta/a_0$ and expanding to leading order in $1/a_0$ with $\chi$ kept constant. From this we see that, for large $\chi$ in LCF, we should rescale $\theta\to\chi^{1/3}\theta$ and then expand the integrand in powers of $1/\chi$. For the other one-step terms we first recall that the LCF approximation is obtained from our expressions in~\cite{Dinu:2018efz} for the exact result by changing variables from $\phi_1, ...,\phi_4$ to $\phi=(\sigma_{43}+\sigma_{21})/2$, $\varphi=\sigma_{43}-\sigma_{21}$, $\theta=(\theta_{43}+\theta_{21})/2$ and $\eta=\theta_{43}-\theta_{21}$, where $\sigma_{ij}=(\phi_i+\phi_j)/2$ and $\theta_{ij}=\phi_i-\phi_j$, and then the one-step terms are obtained by rescaling $\{\varphi,\theta,\eta\}\to\{\varphi,\theta,\eta\}/a_0$ and expanding to leading order in $1/a_0$, which is $\mathcal{O}(a_0)$ (so this is a Laurent series) for the one-step terms (compared to $\mathcal{O}(a_0^2)$ for the two-step). The integrands now only depend on $\phi$ via the locally-constant value of $\chi(\phi)$. So, if we either consider a constant field or the ``rate'' given by the $\phi$ integrand, then we have integrals over $\{\varphi,\theta,\eta\}$ for $\mathbb{P}^{22\to1}$ (the $\varphi$ integral is trivial for $\mathbb{P}^{22\to1}$) and $\mathbb{P}_{\rm ex}^{22}$, for $\mathbb{P}_{\rm dir}^{12}$ and $\mathbb{P}_{\rm ex}^{12}$ we have integrals over $\{\theta,\eta\}$, and for $\mathbb{P}^{11}$ there is only one integral over $\theta$. For $\mathbb{P}^{22\to1}$, $\mathbb{P}_{\rm dir}^{12}$ and $\mathbb{P}_{\rm ex}^{12}$ one can, by looking at the exponential part of the integrand as for $\mathbb{P}^{11}$, see that one should rescale $\{\varphi,\theta,\eta\}\to\chi^{1/3}\{\varphi,\theta,\eta\}$ to obtain the large-$\chi$ expansion. The exponential part of the integrand for $\mathbb{P}_{\rm ex}^{22}$ is more complicated, even in the LCF case. However, it is straightforward to check that the same rescaling of the integration variables also works for $\mathbb{P}_{\rm ex}^{22}$. It turns out that each term has the same form as the total one-step, which we find to be given by
\be\label{PoneLargeExpansion}
\mathbb{P}_{\rm one}(s)=\alpha^2 a_0\Delta\phi\frac{1}{\chi^{1/3}}L \qquad L=\sum_{n=0}^\infty\frac{L_n}{(\chi^{2/3})^n} \;,
\ee  
where the coefficients $L_n$ are obtained by performing the $\{\varphi,\theta,\eta\}$ integrals. We have performed some of these integrals numerically, but this is straightforward and can be done quickly with e.g. Mathematica, even for the exchange term. So, we have without much numerical effort calculated the first 11 terms, i.e. up to $n=10$.
At least from these terms, it seems that the large-$\chi$ expansion is convergent. In any case, just a direct summation of these terms, without any resummation, gives a good precision down to $\chi\sim1$, as can be seen in Fig.~\ref{largechiexpFig}. As $\chi$ decreases below $\chi=1$ the direct sum of this series starts to deviate more and more from the exact result. A quick way to improve this is to make a Pad\'e approximant for $L$ in the variable $y=1/\chi^{2/3}$. Although the (near) diagonal approximants ($[N/N]$, $[(N-1)/N]$ or $[N/(N-1)]$) usually give good improvement, in this case we know that the result should vanish exponentially fast as $\chi\to0$, which means that we find a much better result by using off-diagonal approximants $[M/N]$ with $N\gg M$. In Fig.~\ref{largechiexpFig} we show that $[1/9]$ gives in this case a significant improvement, where, on the scale of this plot, one could argue that we have a decent precision for arbitrary $\chi$. However, if one zooms in on the region where the probability starts to become exponentially suppressed, then one notices that even the Pad\'e-resummed result starts to deviate more and more from the exact result, see Fig.~\ref{MGresumRelErFig}. However, in Fig.~\ref{largechiexpFig} and~\ref{MGresumRelErFig} we see that we obtain a significant improvement with the following, new resummation method, which gives a much higher precision down to much smaller $\chi$.

\subsection{New resummation method}

In this section we will propose a new resummation method, which works very well for these terms. We know that $\mathbb{P}_{\rm one}(s)$ has a large-$\chi$ expansion in the form of~\eqref{PoneLargeExpansion}, and a small-$\chi$ expansion in the form of (note that this starts at $\sqrt{\chi}$ because the leading order terms proportional to $1/\sqrt{\chi}$ in the direct and exchange parts cancel)
\be\label{PoneSmallExpansion}
\mathbb{P}_{\rm one}(s)=\alpha^2 a_0\Delta\phi\sqrt{\chi}\exp\left(-\frac{2r}{3\chi}\right)T 
\ee 
where $r=(1/[1-q_1-q_2]-1)$ and
\be
T=\sum_{n=0}^\infty T_n\chi^n \;.
\ee
The idea now is to look for some special function that has the same type of expansions at both large and small $\chi$. We are inspired by~\cite{Mera:2018qte} to look for such a function starting with a general Meijer-G function~\cite{BatemanBook,LukeBook1969,LukeBook}
\be
G_{pq}^{mn}\left(\begin{matrix} a_1,...,a_p \\ b_1,...,b_q\end{matrix}\bigg|z\right) \;.
\ee
Many special functions can be expressed in terms of $G_{pq}^{mn}$ and having a large number of free parameters ($a_i$ and $b_i$) makes this a very general class of functions and hence a good place to start looking for resummation functions. In~\cite{Mera:2018qte} it was shown how asymptotic series (especially those with branch cuts) can be resummed into a single Meijer-G function on the form $G_{p,p+1}^{p+1,1}$ and where the precision of the resummation is improved by increasing the number of parameters, i.e. increasing $p$. To resum our series in $\chi$ we will end up with a different class of Meijer-G functions. 

For large $\chi$ we want an expansion powers of $1/\chi^{2/3}$, but, since we know that for general photon momenta $q_1$ and $q_2$ the small-$\chi$ expansion has an exponential part as in~\eqref{PoneSmallExpansion}, we use $\xi:=(r/\chi)^{2/3}$ as a rescaled parameter. Apart from the overall factor of $1/\chi^{1/3}\propto\sqrt{\xi}$, the large-$\chi$ expansion should only involve integer powers of $\xi$ as seen in~\eqref{PoneLargeExpansion}. So, we start with $G_{pq}^{mn}(c\xi^k)$, where $k$ is some positive integer and $c$ a constant. At small $\chi$, i.e. large argument of $G_{pq}^{mn}(c\xi^k)$, we want an exponential scaling as in~\eqref{PoneSmallExpansion}. The relevant expansions in this limit can be found in~\cite{BatemanBook,LukeBook1969,LukeBook,FieldsAsymptotic}, and for a general $G_{pq}^{mn}$ this involves first expressing $G_{pq}^{mn}$ as a linear combination of 
$G_{pq}^{q1}$ and $G_{pq}^{q0}$. The latter has the exponential scaling that we want
\be\label{G0exp}
G_{pq}^{q0}(z)\sim\exp\left(-\nu z^{1/\nu}\right)z^\gamma\sum_{n=0}^\infty c_n z^{-n/\nu}\;,
\ee
where $\nu=q-p$ and 
\be\label{gammaDefinition}
\gamma=\frac{1}{\nu}\left(\frac{1-\nu}{2}+\sum_{i=1}^q b_i-\sum_{i=1}^p a_i\right) \;.
\ee
By matching the exponents in~\eqref{PoneSmallExpansion} and~\eqref{G0exp} we find that we need $\nu=2k/3$. Since $\nu$ and $k$ are integers, this implies 
\be
k=3j \qquad \nu=2j \;,
\ee 
where $j=1,2,3...$. This matching also gives
\be
c=9^{-j}j^{-2j}
\ee
and for a function which starts its small-$\chi$ expansion with $\sqrt{\chi}$ rather than $\sqrt{\chi}\chi^n$ we also have
\be
\gamma=-\frac{1}{4j}  \qquad \sum_{i=1}^p a_i=1-j+\sum_{i=1}^q b_i \;.
\ee

As we will demonstrate, we can obtain a good resummation already with only $j=1$, which gives $G_{p,p+2}^{p+2,0}(\xi^3/9)$. The series expansion at $\xi\ll1$ is given by
\be
G_{pq}^{mn}(z)=\sum_{k=1}^m\dots z^{b_k}{}_pF_{q-1}((-)^{p-m-n}z) \;, 
\ee 
where all the parameters have been suppressed and $F$ is the generalized hypergeometric function, which can be expanded in integer powers of $z$. Since the function we are looking for should start with an overall factor of $1/\chi^{1/3}\propto\sqrt{\xi}$, the $b$ parameters can only be
\be
b_k=\frac{1}{6}+\frac{n}{3} \quad n=0,1,2... \;.
\ee 
If two $b$ parameters are equal or differ by an integer, then the expansion of $G$ would involve $\log\chi$ terms, which we do not have here. This implies that we can only have three $b$ parameters, i.e. $q\leq3$, which, since $q=p+2$, means $q=2$ or $q=3$. However, $q=2$ does not work for the following reason: For $j=1$ we have $\gamma=-1/4$ in~\eqref{gammaDefinition}. But with $b_1=1/6+(n_1/3)$ and $b_2=(1/6)+(n_2/3)$ we find $\gamma+1/4=(1+n_1+n_2)/6\ne0$, so there are no $b_1$ and $b_2$ for which $\gamma=-1/4$. So, for this choice of $j=1$, we have $p=1$ and $q=3$. Since $G$ is symmetric with respect to the $b$ parameters, we can without loss of generality set
\be
b_1=\frac{1}{6}+n_1 \quad b_2=\frac{1}{6}+\frac{1}{3}+n_2 \quad b_3=\frac{1}{6}+\frac{2}{3}+n_3
\ee      
and then $\gamma=-1/4$ implies
\be
a_1=\frac{3}{2}+n_1+n_2+n_3+n \;,
\ee 
where $n=0,1,2...$.
Note that $n>0$ corresponds to functions with expansions starting at a higher order, i.e. $\sqrt{\chi}\chi^n$ rather than $\sqrt{\chi}$, but these are relevant because our resummation involves a sum of different $G$ functions. However, the different choices of $n$ and $n_i$ do not all give independent functions: It is easy to show using the Mellin-Barnes integral definition of $G$ and $\Gamma(z+1)=z\Gamma(z)$ that these functions obey the following contiguous relations
\be
G(n,n_3)=G(n,n_3-1)+(b_3-a_1)G(n+1,n_3-1) 
\ee  
and similarly for $n_1$ and $n_2$. This means that any function with nonzero $n_i$ can be reduced to a linear combination of functions with different $n$ and $n_i=0$, so we can without loss of generality set $n_1=n_2=n_3=0$. Thus, we have finally found a set of resummation functions,
\be
f_n(\chi):=\frac{\sqrt{3}\Gamma\left[\frac{4}{3}+n\right]}{2\pi}\sqrt{\xi}G_{13}^{30}\left(\begin{matrix} \frac{4}{3}+n \\ 0,\frac{1}{3},\frac{2}{3}\end{matrix}\bigg|\frac{\xi^3}{9}\right) \;,
\ee
where the overall normalization constant is chosen such that the large-$\chi$ expansion starts with $f_n=1+\mathcal{O}(\xi)$. The factor of $\sqrt{\xi}$ in the prefactor comes from using
\be
G\left(\begin{matrix} a_i+c \\ b_j+c\end{matrix}\bigg|z\right)=z^c G\left(\begin{matrix} a_i\\ b_j\end{matrix}\bigg|z\right) \;,
\ee
with $c=1/6$.
The resummation is now obtained by matching the original series onto the corresponding expansion of 
\be\label{sumofG}
\sum_{n=0}^N c_n f_n(\chi) \;,
\ee
which determines the constants $c_n$. 

Although expansion formulas for general $G_{pq}^{mn}$ can be found in~\cite{BatemanBook,LukeBook1969,LukeBook,FieldsAsymptotic}, especially the higher-order terms in the small-$\chi$ expansions can be difficult to find. So, we will for convenience explain how to obtain the expansions for the particular $G$ function that we have. The large-$\chi$ expansion can be found starting with the Mellin-Barnes integral representation, which in our case gives
\be
\begin{split}
f_n=&\frac{\sqrt{3}\Gamma\left(\frac{4}{3}+n\right)}{2\pi}\sqrt{\xi}\int_{-i\infty}^{i\infty}\frac{\ud s}{2\pi i}\left(\frac{\xi^3}{9}\right)^s \\
&\hspace{1cm}\times\frac{\Gamma(-s)\Gamma\left(\frac{1}{3}-s\right)\Gamma\left(\frac{2}{3}-s\right)}{\Gamma\left(\frac{4}{3}+n-s\right)} \\
=&\Gamma\left(\frac{4}{3}+n\right)\int_{-i\infty}^{i\infty}\frac{\ud s}{2\pi i}\frac{3^{1+s}\xi^{\frac{1}{2}+3s}\Gamma(-3s)}{\Gamma\left(\frac{4}{3}+n-s\right)} \;,
\end{split}
\ee  
where we have used Gauss's multiplication formula for $\Gamma$~\cite{DLMF}. 
The $s$ integral has poles at $s=m$, $s=(1/3)+m$ and $s=(2/3)+m$, where $m=0,1,2,...$, and the expansion in powers of $\xi$ can now be obtained by performing this integral with Cauchy's residue theorem.

As an aside, we note that, while a Meijer-G function can always, trivially be expressed as a Fox-H function~\cite{FoxHbook} with twice as many parameters, from the above equation we see that $f_n$ can actually be expressed compactly as a different Fox-H function,
\be
f_n=3\Gamma\left(\frac{4}{3}+n\right)\sqrt{\xi}H_{11}^{10}\left(\begin{matrix} \left(\frac{4}{3}+n,1\right) \\[.2cm] \left(0,3\right)\end{matrix}\bigg|3\xi^3\right) \;,
\ee
A similar reformulation could be more useful in the generalization to Meijer-G functions with more parameters, i.e. to $j>1$.

Note that, using this integral representation, the sum in~\eqref{sumofG} can be expressed as a single Mellin-Barnes integral, so this resummation gives an approximation of the Mellin transform of the probability with respect to $\chi$. It could be interesting to study whether one could find a resummation directly in terms of the Mellin transform rather than finding one via $G$.  

To obtain the small-$\chi$ expansion, it is convenient to rewrite the Mellin-Barnes integral as a ``LCF''-type integral: We obtain this by first using~\cite{DLMF}
\be
\Gamma(-3s)=\int_0^\infty\frac{\ud t}{t}e^{-t}t^{-3s}
\ee
and
\be
\frac{1}{\Gamma(a_1-s)}=\int\frac{\ud u}{2\pi i}e^u u^{-a_1+s} \;,
\ee
where the integration contour for the $u$ integral goes around the negative axis counterclockwise. We will perform the $s$ and $t$ integrals. In the $s$ integrand we have
\be
\exp\left\{s\left(\ln y-3[\ln|t|+i\text{arg}(t)]+\ln|u|+i\text{arg}(u)\right)\right\} \;,
\ee
where $y=3\xi^3$.
To simplify this we choose an integration contour for the $t$ integral with $\text{arg}(t)=(1/3)\text{arg}(u)$ and change variable from $t=e^{T+i\text{arg}(u)/3}$ to $T$, where $T$ goes from $-\infty$ to $\infty$ on the real axis. Then the $s$ integral gives a delta function, which we use to perform the $T$ integral. We obtain
\be\label{fnLCFrepresentation}
\begin{split}
f_n(\chi)=&\Gamma(a_1)\sqrt{\xi}\int\frac{\ud u}{2\pi i}\frac{1}{u^{a_1}}\exp\left\{-(yu)^{1/3}+u\right\} \\
=&-3^{4/3}\left(\frac{3x}{i}\right)^n\Gamma\left(\frac{4}{3}+n\right) \\
&\times\int\frac{\ud\tau}{2\pi}\frac{1}{\tau^{2+3n}}\exp\left\{\frac{i}{x}\left(\tau+\frac{\tau^3}{3}\right)\right\} \;,
\end{split}
\ee 
where $x=1/\xi^{3/2}=\chi/r$ and the $\tau$ contour lies in the upper complex plane (e.g. from $\tau=\infty e^{5i\pi/6}$ to $\tau=\infty e^{i\pi/6}$), or for integration along the real axis the pole at $\tau=0$ is avoided with $\tau\to\tau+i\epsilon$, $\epsilon>0$. The small-$\chi$ expansion is now readily obtained with the saddle-point method by changing variable from $\tau=i+\sqrt{x}\delta\tau$ to $\delta\tau$ and expanding the integrand in $x$. 

The integral in~\eqref{fnLCFrepresentation} is of the type that one usually encounters in LCF. By making partial integration one can rewrite it as
\be
f_n=P_1\left(\frac{1}{x}\right)\text{Ai}_1(\xi)+P_2\left(\frac{1}{x}\right)\frac{\text{Ai}(\xi)}{\sqrt{\xi}}+P_3\left(\frac{1}{x}\right)\frac{\text{Ai}'(\xi)}{\xi} \;,
\ee 
where $\text{Ai}(\xi)$ is the Airy function,
\be
\text{Ai}_1(\xi)=\int_\xi^\infty\ud t \text{Ai}(t) \;,
\ee
and $P_i$ are polynomials. This might seem like a simpler formulation of this resummation. However, these polynomials are not arbitrary, but related by the fact that the small-$\chi$ expansions should not have negative powers of $\chi$.
In any case, we see that for some terms in LCF this resummation will converge to the exact result after summing a finite number of $f_n$. For example, here we find that the one-step contribution from $\mathbb{P}^{11}$ is exactly given by $f_0(\chi)$, so if we resum $\mathbb{P}^{11}$ separately then we simply have $c_n=0$ for all $n>0$ in~\eqref{sumofG}. This is perhaps not so surprising given that exact results for $\mathcal{O}(\alpha)$ processes have been expressed in terms of Meijer-G functions in~\cite{Lobanov1980}, and $\mathbb{P}^{11}$ has a structure similar to $\mathcal{O}(\alpha)$ processes. 

In principle, one could use both the small-$\chi$ and large-$\chi$ expansions of $\mathbb{P}$ to determine the coefficients $c_n$ in~\eqref{sumofG}. However, in this case it turns out to be much better to use more terms from the large-$\chi$ expansion. In fact, we find a very good resummation using only the large-$\chi$ expansion. For example, in Fig.~\ref{largechiexpFig} we use only the large-$\chi$ expansion and find a resummation that works down to much smaller $\chi$ compared to the Pad\'e resummation, even using only half as many terms from~\eqref{PoneLargeExpansion}. The reason for this is that the new resummation has the same exponential scaling at small $\chi$ as the exact result, while the Pad\'e approximant can only approximate this using a large power. So, for a fixed order the Pad\'e approximant will break down as $\chi$ decreases, while the new resummation can still give a good approximation at small $\chi$, even if none of the coefficients $c_n$ in~\eqref{sumofG} are determined with the coefficients $T_n$ in~\eqref{PoneSmallExpansion}. In Fig.~\ref{MGresumRelErFig} we show that the relative error can be made very small, even at small $\chi$, by including more terms in~\eqref{sumofG}. 
Having a resummation of the large-$\chi$ expansion that works down to such small $\chi$ is very useful, because this means that there will be a significant overlap with the resummations of the small-$\chi$ expansion, even for a simple resummation such as Borel $+$ Pad\'e, which in turn means that one can check the precision of these resummations without using any numerical data for the exact result. Using the new resummation to resum the large-$\chi$ expansion we find results that, for a given order, eventually starts to deviate from the exact result at small $\chi$ as seen in Fig.~\ref{MGresumRelErFig}. However, the result vanishes exponentially as $\chi\to0$, so it is not necessarily very useful to have a high precision at very small $\chi$ anyway.

\subsection{LCF integrals}

Applying the LCF approach to some non-constant field, means replacing $\chi$ in the above with the locally constant $\chi(\sigma)$ and integrating over $\sigma$. (In doing this one should replace the overall factor of $a_0$ in e.g.~\eqref{PoneSmallExpansion} with $\chi(\sigma)/b_0$.)
Here it is an advantage that the above resummation methods give the result for an entire interval in $\chi$ rather than just the result for a single value of $\chi$. In other words, the output of one resummation is a function of $\chi$, not just a number. We just have to make sure that the resummation function is valid up to the maximum of $\chi(\sigma)$ and down to values of $\chi$ where the $\sigma$ integrand starts to become negligible. Since the dependence on $\chi$ is slow, the integral over $\sigma$ is not difficult to perform numerically. For some field shapes we can even perform the $\sigma$ integral analytically: 

Consider for example $a'(\sigma)=a_0\sin^2\sigma$ for $0<\sigma<\pi$, i.e. $\chi(\sigma)=\chi_0\sin^2\sigma$ where $\chi_0=a_0b_0$. We are motivated to consider such a short pulse since this makes the one-step terms more important compared to the two-step. (We could trivially consider a train of such pulses, e.g. with different sign such that $a(\infty)=a(-\infty)$.) Using the Mellin-Barnes integral representation of $G$ and
\be
\int_0^\pi\ud\sigma\sin^c\sigma=\frac{\sqrt{\pi}\Gamma\left[\frac{1+c}{2}\right]}{\Gamma\left[1+\frac{c}{2}\right]} \;,
\ee
which follows from a suitable integral representation of the Beta function $\Gamma(a)\Gamma(b)/\Gamma(a+b)$ (see~\cite{DLMF}),
we find
\be
\begin{split}
h_n(\chi_0):=&\int\ud\sigma\,\chi(\sigma)f_n[\chi(\sigma)] \\
=&\frac{\sqrt{3}\Gamma\left(\frac{4}{3}+n\right)}{2\sqrt{2\pi}}\sqrt{\xi_0}G_{35}^{50}\left(\begin{matrix} \frac{5}{6},\frac{4}{3},\frac{4}{3}+n \\[.1cm] 0,\frac{1}{3},\frac{2}{3},\frac{7}{12},\frac{13}{12}\end{matrix}\bigg|\frac{\xi_0^3}{9}\right) \;,
\end{split}
\ee 
where $\xi_0=(r/\chi_0)^{2/3}$. So, having found the coefficients $c_n$, going from a constant field to this pulsed field is simply done by replacing $f_n$ in~\eqref{sumofG} with $h_n$.
 
For an oscillating field, $a'(\sigma)=a_0\sin\sigma$ and $\chi(\sigma)=\chi_0|\sin\sigma|$, we find a similar result for the integral over each cycle
\be
\begin{split}
&\int_0^\pi\ud\sigma\,\chi(\sigma)f_n[\chi(\sigma)] \\
&=\frac{\sqrt{3}\Gamma\left(\frac{4}{3}+n\right)}{2\sqrt{\pi}}\sqrt{\xi_0}G_{24}^{40}\left(\begin{matrix} \frac{4}{3},\frac{4}{3}+n \\[.1cm] 0,\frac{1}{3},\frac{2}{3},\frac{5}{6}\end{matrix}\bigg|\frac{\xi_0^3}{9}\right) \;.
\end{split}
\ee 

For a Gaussian pulse, $a'(\sigma)=a_0 e^{-\sigma^2}$, the $\sigma$ integral gives
\be
\int\ud\sigma\exp\left\{-\left(\frac{2}{3}-2s\right)\sigma^2\right\}=\frac{\sqrt{3\pi}}{\sqrt{2-6s}} \;.
\ee
The square root means that the $s$ integrand now has a branch cut and can therefore not be performed in terms of a Meijer-G function. However, the result is still expressed as a single Mellin transform, and a common way of evaluating Meijer-G functions is anyway to perform a 1D integral, so it is not really a problem if a field shape leads to a Mellin transform that cannot be expressed as a Meijer-G function. 

Thus, having expressed the $\chi$ dependence of the probability as a Mellin transform can be quite useful for going from a constant to an inhomogeneous field in LCF.

\section{Resummation of small- and large-$\chi$ expansions}\label{resumSmallLargeSection}

In this section we will show how to simultaneously resum both the small- and the large-$\chi$ expansions. In the previous section we resummed into a sum of single Meijer-G functions~\eqref{sumofG}, where the difference between the terms is governed by an integer. We saw that this works well for the resummation of the large-$\chi$ expansion, but less well for the small-$\chi$ expansion. One could view this as the resummation functions being too rigid for the small-$\chi$ expansion. In this section we will therefore present another, more flexible resummation method. We are inspired here by the resummation method in~\cite{Alvarez:2017sza}, which we used in Sec.~\ref{resumSmallSection}, to resum the small-$\chi$ expansion with the exponential part factored out, i.e. for $T$ in~\eqref{PoneSmallExpansion}. In this section we are interested in something similar to~\eqref{ASresumDefinition}, but for the whole probability, not just the exponential part. One can expect that treating the whole probability could be useful and allow for an improved resummation, e.g. because if one factors out the exponential $\exp(-2r/[3\chi])$, then the large-$\chi$ expansion of it gives a power series in $1/\chi$, which would have to be compensated somehow by the expansion of the resummed $T$ in order to obtain the correct series for the whole probability, which we know only involves factors of $1/\chi^{2/3}$. It therefore seems advantageous to look for resummations of the whole rather just part of the probability. However, we see immediately that we cannot simply take one of the Meijer-G functions from the previous section and rescale its argument, because if $G(\chi)$ gives the correct exponential as in~\eqref{PoneSmallExpansion}, then $G(\chi/\chi_0)$ gives a different exponential and hence a very different scaling. This is in contrast to a resummation of a series without exponential part, because if $\phi(\chi)$ has the correct type of power series expansion, so too does $\phi(\chi/\chi_0)$. To overcome this obstacle, we propose a resummation which is quadratic in (for example) Meijer-G functions. 

So, we are looking for a function $f(\chi/r)$ that, roughly speaking, has the same type of small- and large-$\chi$ expansions as the square root of the probability (or rather the one-step part). Using arguments similar to the previous section, we find that one such function is given by
\be\label{fforGG}
f(x)=-\frac{\Gamma\left[-\frac{1}{12}\right]}{8\sqrt{3}\pi x^{1/6}}G_{13}^{30}\left(\begin{matrix} \frac{11}{12} \\[.1cm] 0,\frac{1}{3},\frac{2}{3}\end{matrix}\bigg|\frac{1}{9x^2}\right) \;.
\ee
Then the following function has the same type of small- and large-$\chi$ expansions as the probability,
\be\label{doubleGdefinition}
F(w;x)=(1-w^2)^{1/4}f\left(\frac{2x}{1+w}\right)f\left(\frac{2x}{1-w}\right) \;,
\ee
where $w$ is a continous parameter. Note that for $w\ne0$ the two separate factors of $f$ each has a different exponential scaling than $\mathbb{P}_{\rm one}$, but their product still has the same exponetial scaling as $\mathbb{P}_{\rm one}$ because $(1+w)/2+(1-w)/2=1$. The factor of $(1-w^2)^{1/4}$ in the prefactor is just for convenience. We also find it convenient to change variable from $w=\sqrt{1-(1/\nu)}$ to $\nu$. The resummation is now given by
\be\label{newQuadResum}
\sum_{n=1}^N c_n F\left(w_n;\frac{\chi}{r}\right) \;,
\ee
where the constants $c_n$ and $w_n$ (or $\nu_n$) are obtained by matching with the small- and large-$\chi$ expansions of $\mathbb{P}$. In~\cite{Alvarez:2017sza} it was shown that the coefficients in~\eqref{ASresumDefinition} can be obtained conveniently from a Pad\'e approximant. What we have here is more complicated and we cannot use the same method. Instead we have simply obtained $c_n$ and $w_n$ by a numerical root-finding with the Newton-Raphson method and just guessing a starting point. Fortunately, this is not a big problem because, as we will show, we do not actually need to include many terms. 
The constants $c_n$ and $w_n$ can be obtained by using only the small-$\chi$ expansion, but using terms from both expansions allows us to find a resummation that converges to the exact result as both $\chi\to0$ and $\chi\to\infty$. So, in some cases this will give a uniform resummation with a maximum relative error at some finite $\chi$. In the particular case considered here, $\mathbb{P}_{\rm one}$ changes sign at $\chi\sim3$, which means that there will be a short interval around this point where the relative error diverges. However, this is not a real problem since it is just due to the fact that the exact result goes to zero, and in effect we have a resummation which gives a good precision for any value of $\chi$. 

\begin{figure}
\includegraphics[width=\linewidth]{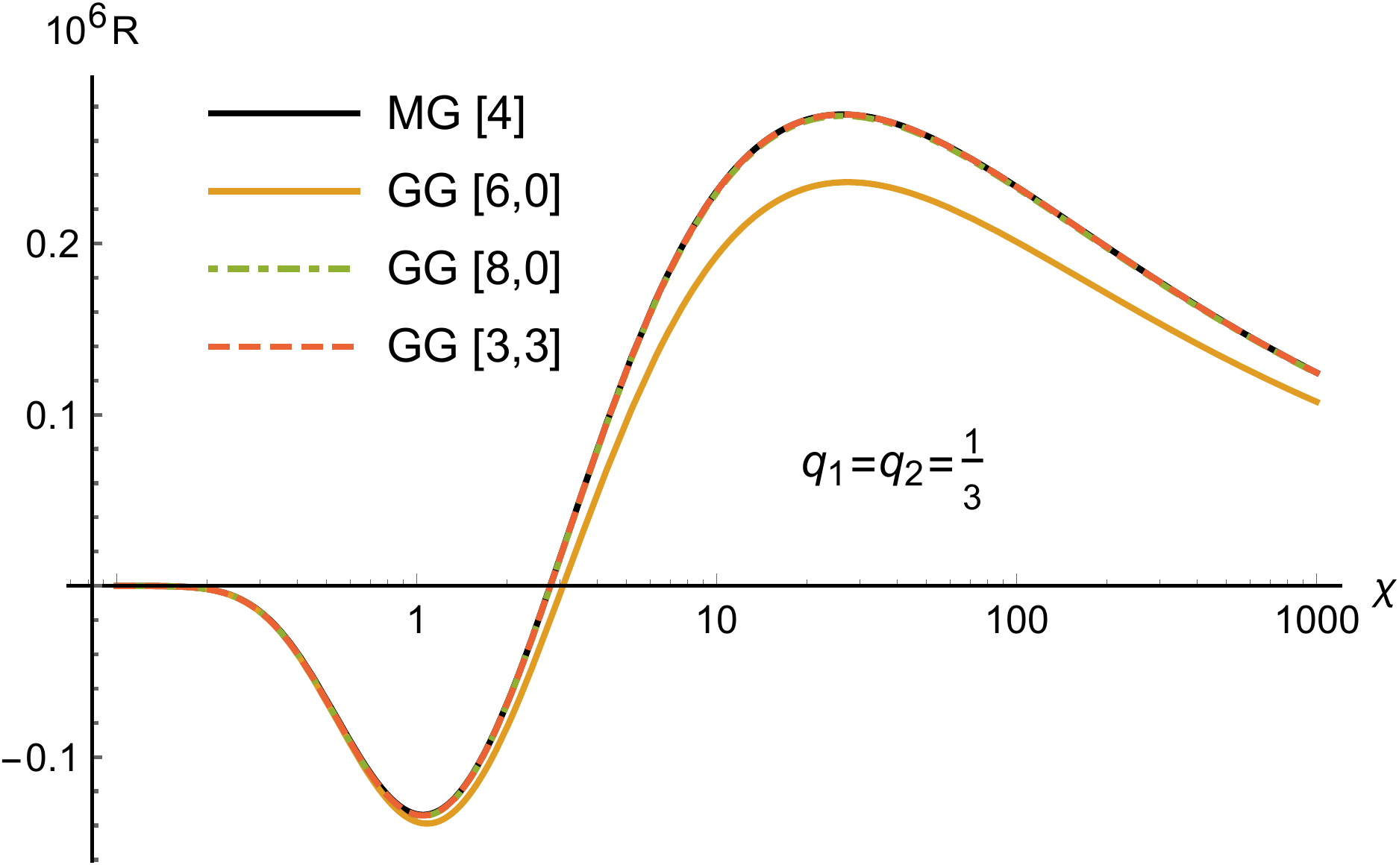}
\caption{Resummation in~\eqref{newQuadResum}. $[m,n]$ means that the first $m$ terms from the small-$\chi$ expansion and the first $n$ terms from the large-$\chi$ expansion have been used to determine the constants $c_n$ and $w_n$ in~\eqref{newQuadResum}. The MG line is the linear Meijer-G resummation~\eqref{sumofG}, which, as demonstrated in Fig.~\ref{largechiexpFig}, effectively represents the exact result on the scale of this plot.}
\label{GGfig}
\end{figure}

As an example, we consider $q_1=q_2=1/3$. With $N=3$ in~\eqref{newQuadResum}, we find by matching with the first 3 terms in the small- and the large-$\chi$ expansions: $c_1\sim1.3-0.26i$; $\nu_1\sim1.1+0.66i$; $c_2=c_1^*$; $\nu_2=\nu_2^*$; $c_3\sim-2.9$; $\nu_3\sim2.0$. We see that these constants are in general complex, but the complex constants come in conjugate pairs, ensuring that the result is real. The relative error is less than $0.1\%$ for $\chi\lesssim1.27$ and $\chi\gtrsim3.59$, and becomes increasingly more precise as $\chi$ becomes smaller or larger. The relative error is larger around the point where the result changes sign, but the resummation is nevertheless indistinguishable from the exact result on a plot like Fig.~\ref{largechiexpFig}. So, with only 3 terms each from the small- and the large-$\chi$ expansions, we obtain a resummation that works for any $\chi$. This is illustrated in Fig.~\ref{GGfig}.

If we only use the small-$\chi$ expansion, then we need to go to $N=4$ in order to have a precise resummation at large $\chi$, see Fig.~\ref{GGfig}. In this case we find $c_1\sim62.9-98.7i$; $\nu_1\sim1.3+0.075i$; $c_2=c_1^*$; $\nu_2=\nu_2^*$; $c_3\sim0.39$; $\nu_3\sim0.61$; $c_4\sim-126.5$; $\nu_4\sim1.4$. From this we see that in some cases $0<\nu_i<1$, which makes $w_i$ purely imaginary. The result is still real though because~\eqref{doubleGdefinition} is an even function of $w$. 
The relative error is less than $0.3\%$ in the large-$\chi$ limit, which can be seen by noting that the relative error in the first couple of terms in the large-$\chi$ expansion is $\{|(L_1^{\rm re}/L_1)-1|,|(L_2^{\rm re}/L_2)-1|,...\}\sim\{0,0029,0.0028,0.004,0.01,0.007,...\}$, where $L_n$ are the exact expansion coefficients in~\eqref{PoneLargeExpansion} and $L_n^{\rm re}$ are the corresponding ones obtained by expanding~\eqref{newQuadResum}. From this we see that, with the first $8$ terms from the small-$\chi$ expansion, the resummation in~\eqref{newQuadResum} actually gives a good approximation of the first couple of large-$\chi$-expansion coefficients, even though none of the constants in~\eqref{newQuadResum} were obtained by matching with the large-$\chi$ expansion coefficients (i.e. only using the basic fact that the expansion is in powers of $1/\chi^{2/3}$). The fact that $|(L_i^{\rm re}/L_i)-1|$ is already small means that, if we want a higher precision at large $\chi$, then the solution for $c_i$ and $\nu_i$ that was obtained using only small-$\chi$ coefficients also serves as a good starting point for numerically finding the corresponding solution if e.g. $6$ terms are obtained from the small-$\chi$ coefficients and the remaining $2$ from the large-$\chi$ coefficients.

One can imagine many different resummuations that, like~\eqref{fforGG} and~\eqref{doubleGdefinition}, involve products of two or more functions with rescaled arguments such that the exponential part at small $\chi$ remains fixed. For example, with 
\be
f(x)=\frac{2\sqrt{\pi}}{5^{1/6}x^{1/15}}\text{Ai}\left[\frac{1}{(5x)^{2/3}}\right]
\ee 
we have that $f^5(\chi/r)$ has the correct type of small- and large-$\chi$ expansions. So, we can use this function for resummation by matching onto for example
\be\label{Ai5ReVers1}
\sum_{n=1}^N c_n (1-w_n^2)^{1/5}f(x)f^2\left[\frac{x}{1-w_n}\right]f^2\left[\frac{x}{1+w_n}\right]
\ee
or
\be
\begin{split}
\sum_{n=1}^N &c_n(1-v_n^2)^{1/10}(1-w_n^2)^{1/10}f(x) \\
\times&f\left[\frac{x}{1-v_n}\right]f\left[\frac{x}{1+v_n}\right]f\left[\frac{x}{1-w_n}\right]f\left[\frac{x}{1+w_n}\right] \;,
\end{split}
\ee
where $x=\chi/r$ and $c_n$, $v_n$ and $w_n$ are constants to be obtained by matching with the small- and/or large-$\chi$ expansions. These two examples seem to lead to somewhat slower convergence if only the small-$\chi$ expansion coefficients are used. However, with $N=3$ and with, say, three of the first terms from the large-$\chi$ expansion, we again obtain resummations that are indistinguishable from the exact result on a plot like Fig.~\ref{GGfig}. An advantage of resummations that only involve well-used functions like the Airy function is that they can be faster or more convenient to evaluate numerically. 

\begin{figure}
\includegraphics[width=\linewidth]{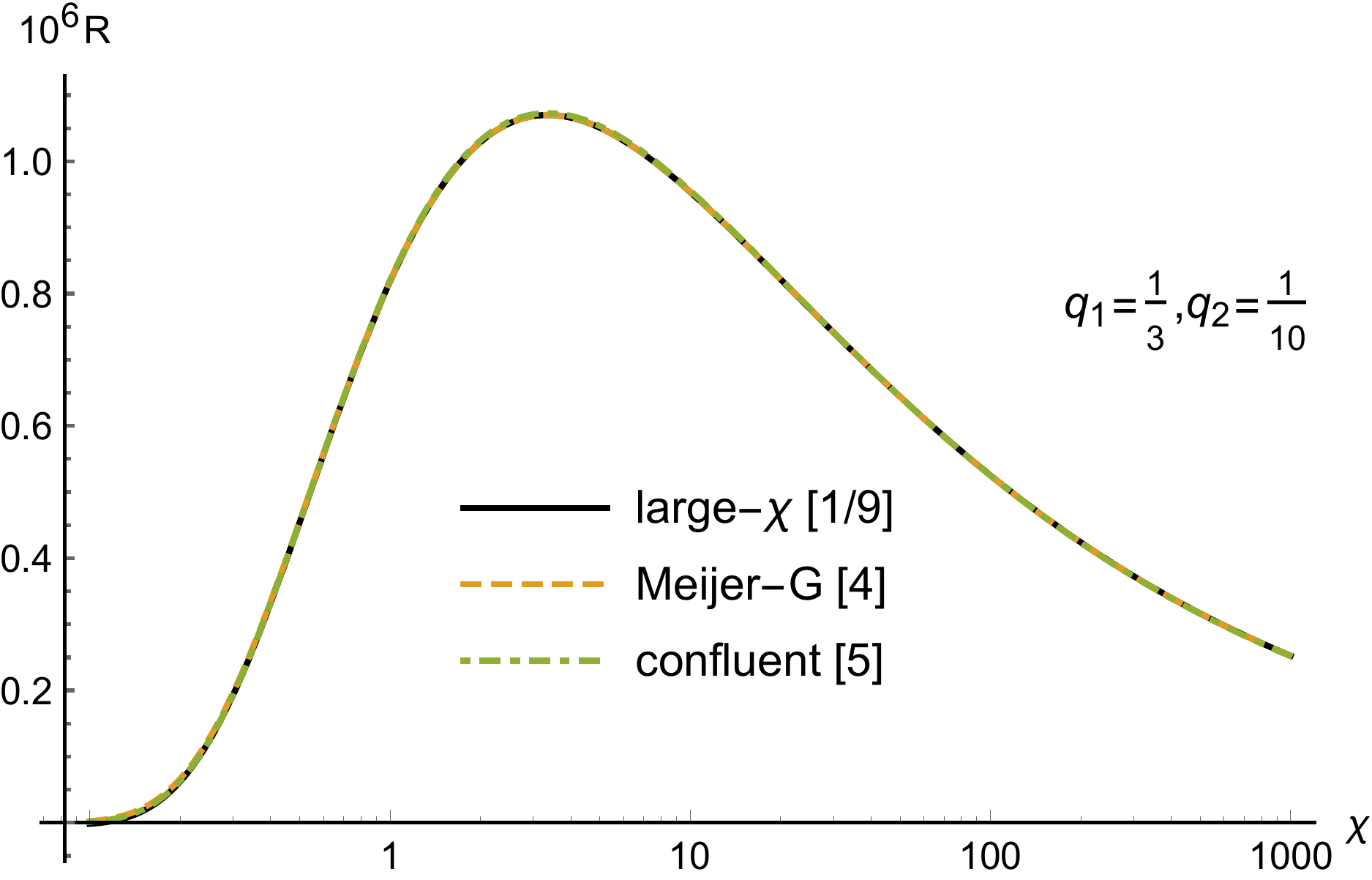}
\includegraphics[width=\linewidth]{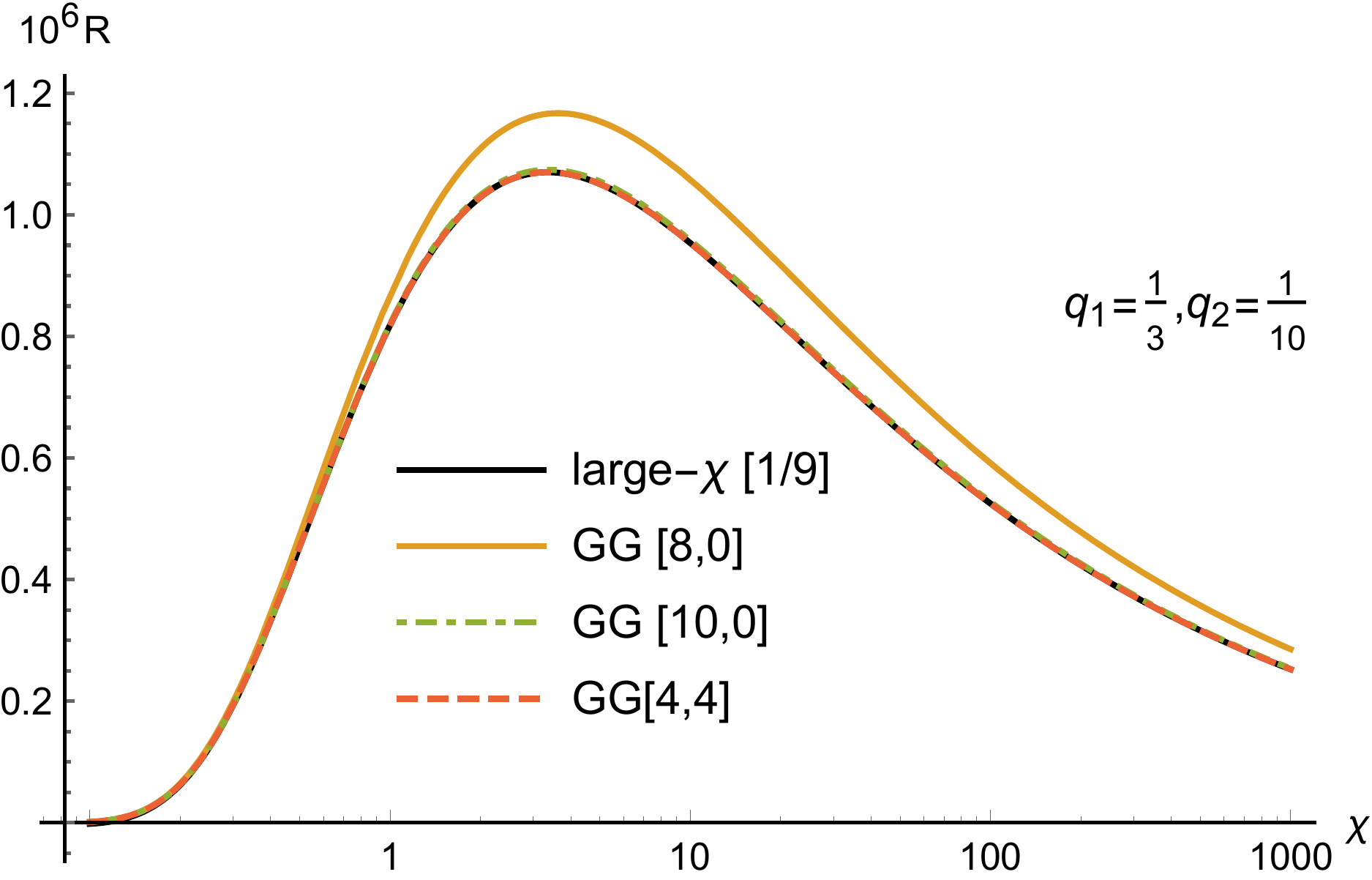}
\caption{Total one-step. Another illustration of the fact that several different small- and/or large-$\chi$ resummations have large overlap. The confluent hypergeometric resummation is of the same type as in Fig.~\ref{largechiexpFig}, i.e. with $n=5$, $a=5/6$ and $b=156/100$.}
\label{overlapFig2}
\end{figure}

\begin{figure}
\includegraphics[width=\linewidth]{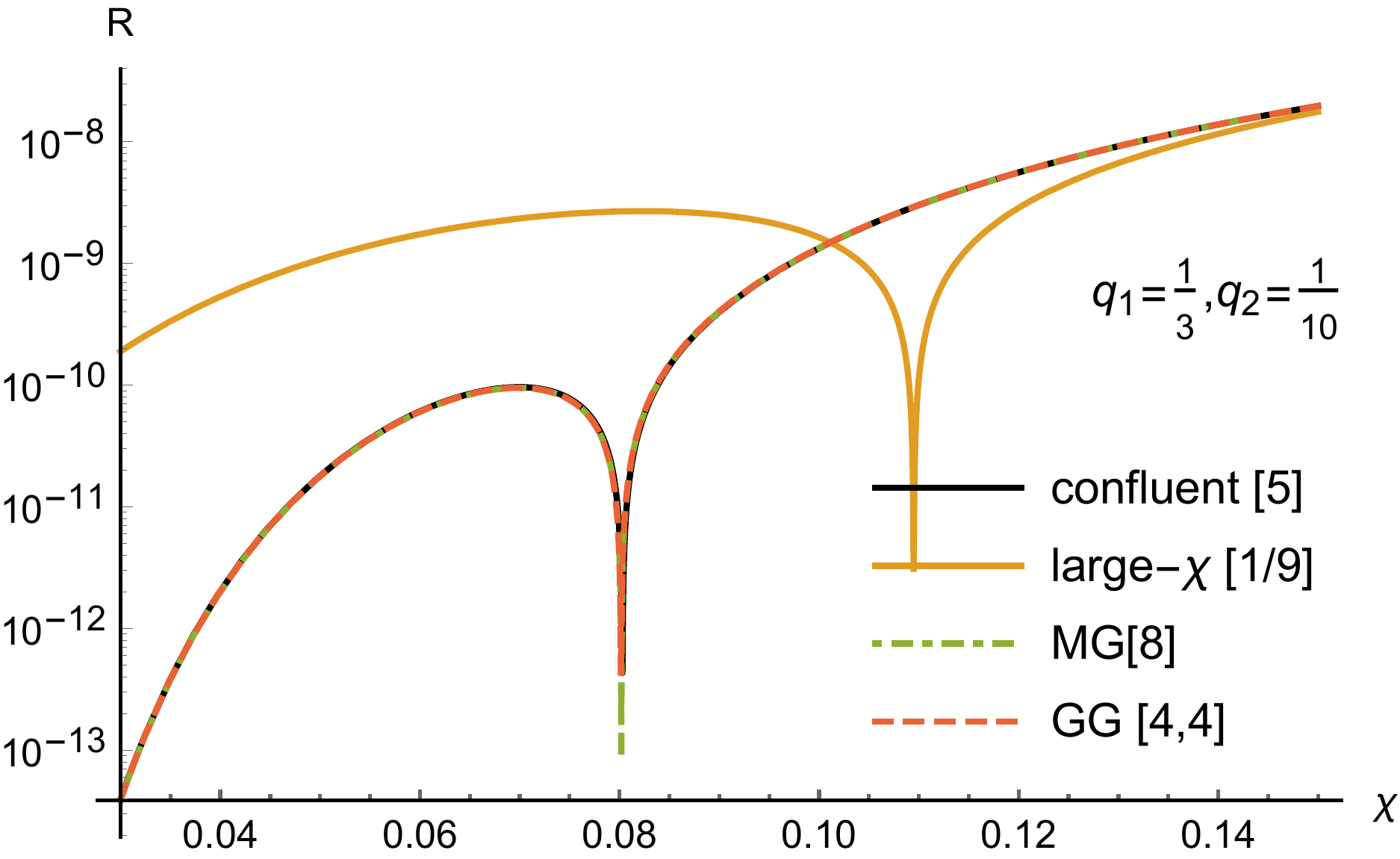}
\caption{Same as Fig.~\ref{overlapFig2}.}
\label{overlapFig2zoom}
\end{figure}

\subsection{Another example}

So far we have used $q_1=q_2=1/3$ as an example. We can of course use the same methods for other points in the longitudinal momentum spectrum. As another example, we consider $q_1=1/3$ and $q_2=1/10$. In Fig.~\ref{overlapFig2} we show that we have several different resummation that give essentially the same result over a large interval of $\chi$. The resummations that use more of the large (small)-$\chi$ expansion coefficients tend to be more precise at larger (smaller) $\chi$. 

For the confluent hypergeometric resummation, note that, while $a$ is fixed by the leading large-$\chi$ scaling, which is $1/\chi^{2/3}$ for any point in the spectrum, $b$ is not fixed by this scaling. In Fig.~\ref{largechiexpFig} (for $q_1=q_2=1/3$) we chose a $b$ that leads to a resummation with good agreement with the large-$\chi$ expansion at large $\chi$, and in Fig.~\ref{overlapFig2} we can see that the same choice of $b$ also gives a good agreement with the other resummations for this second example, $q_1=1/3$ and $q_2=1/10$.  

The simple Pad\'e approximant of the large-$\chi$ expansion has a good agreement with the other resummations on the scale of Fig.~\ref{overlapFig2}. However, by zooming in on smaller $\chi$ in Fig.~\ref{overlapFig2zoom}, we can see that the Pad\'e approximant eventually breaks down as $\chi$ decreases. In contrast, the Meijer-G resummation in~\eqref{sumofG}, using only large-$\chi$ coefficients, is still good even as the result becomes very small.     

\begin{figure}
\includegraphics[width=\linewidth]{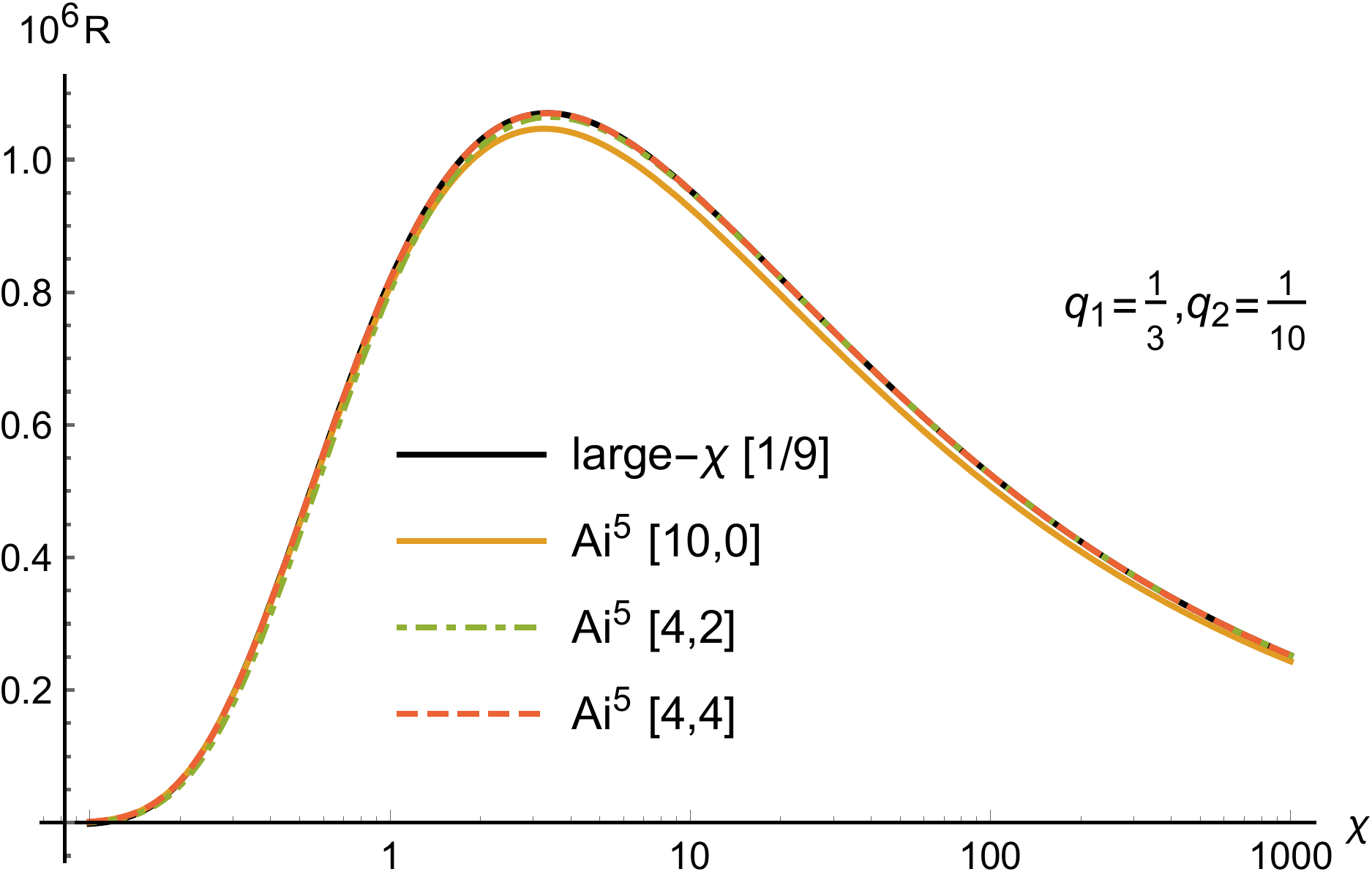}
\includegraphics[width=\linewidth]{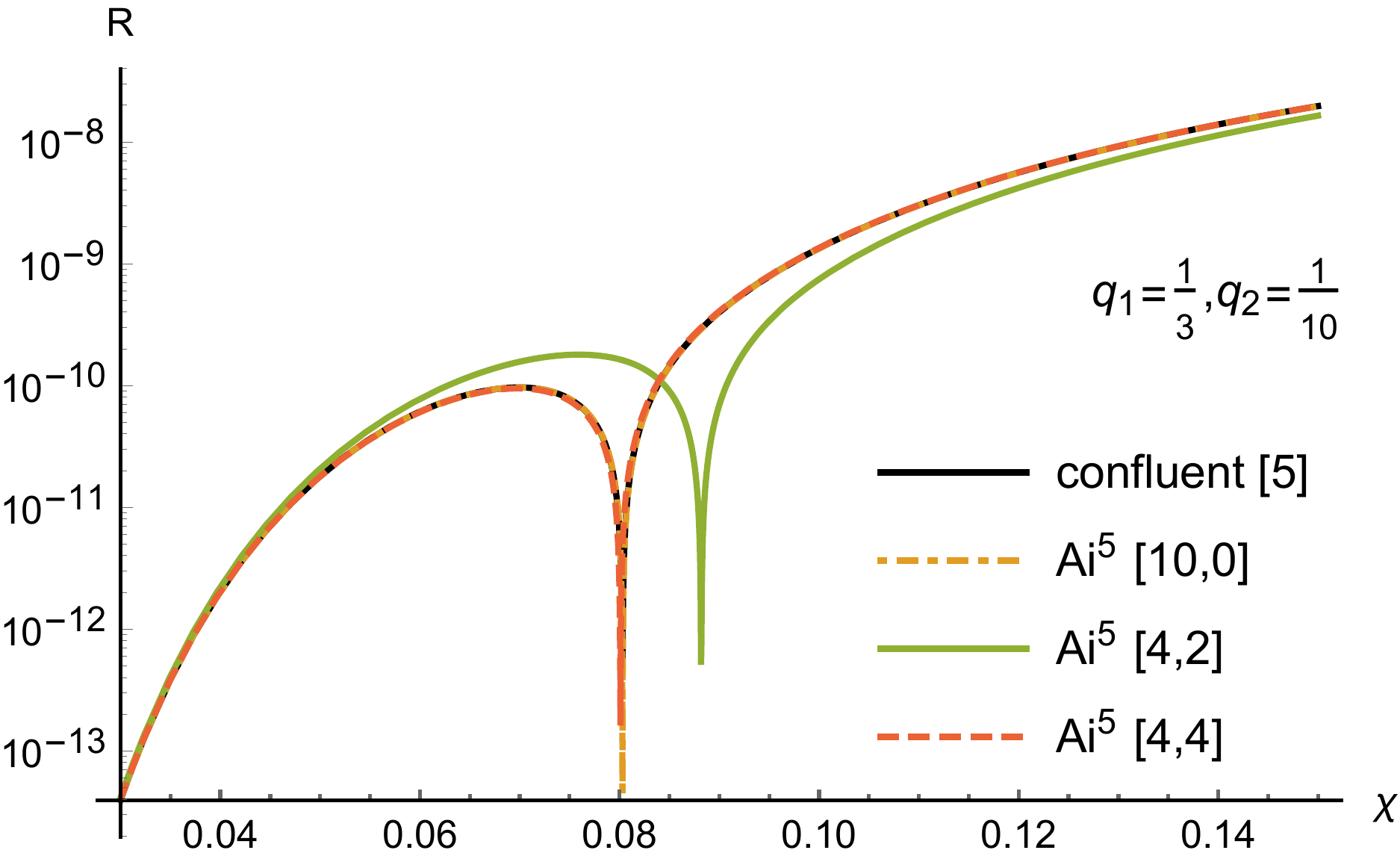}
\caption{Different orders of the Airy-function resummation in~\eqref{Ai5ReVers1}, where $[m,n]$ means that the constants in~\eqref{Ai5ReVers1} are determined by the first $(m,n)$ terms in the (small,large)-$\chi$ expansions.}
\label{AiRe2Fig}
\end{figure}
 
In Fig.~\ref{AiRe2Fig} we consider the Airy-function resummation in~\eqref{Ai5ReVers1}. As for the previous example, if only the small-$\chi$ coefficients are used then this resummation has a somewhat slower convergence compared the Meijer-G resummation in~\eqref{newQuadResum}. However, if we use both the small- and the large-$\chi$ coefficients then we obtain a competitive resummation. With $(4,2)$ terms from the (small,large)-$\chi$ expansions we find a decent precision at $\chi\gtrsim1$, but a significant difference around the point in the small-$\chi$ region where the result changes sign. By matching with two more terms in the large-$\chi$ expansion, i.e. going to $(4,4)$, the precision is naturally increased at $\chi\gtrsim1$, but, more importantly, we obtain a significant improvement at smaller $\chi$, with now a good agreement with the other resummations. So, including more terms from the large-$\chi$ expansion helps also at smaller $\chi$, even though the same number of terms from the small-$\chi$ expansion were used. Thus, with only four terms from the small- and large-$\chi$ expansion, respectively, we obtain a good resummation for arbitrary $\chi$. As mentioned, this Airy-function resummation can be more convenient for numerical evaluation.

\section{Resumming one-step in photon trident}\label{photonTridentResum}

In the previous sections we have showed how to resum the one-step part in double Compton. Exactly the same resummation methods can also be used for photon trident. Here we do not have any previous results to compare with, but, as shown in the previous sections, we have several resummations that can be compared with each other and we find a large overlap for the resummation of the small- and the large-$\chi$ expansions. 

\begin{figure}
\includegraphics[width=\linewidth]{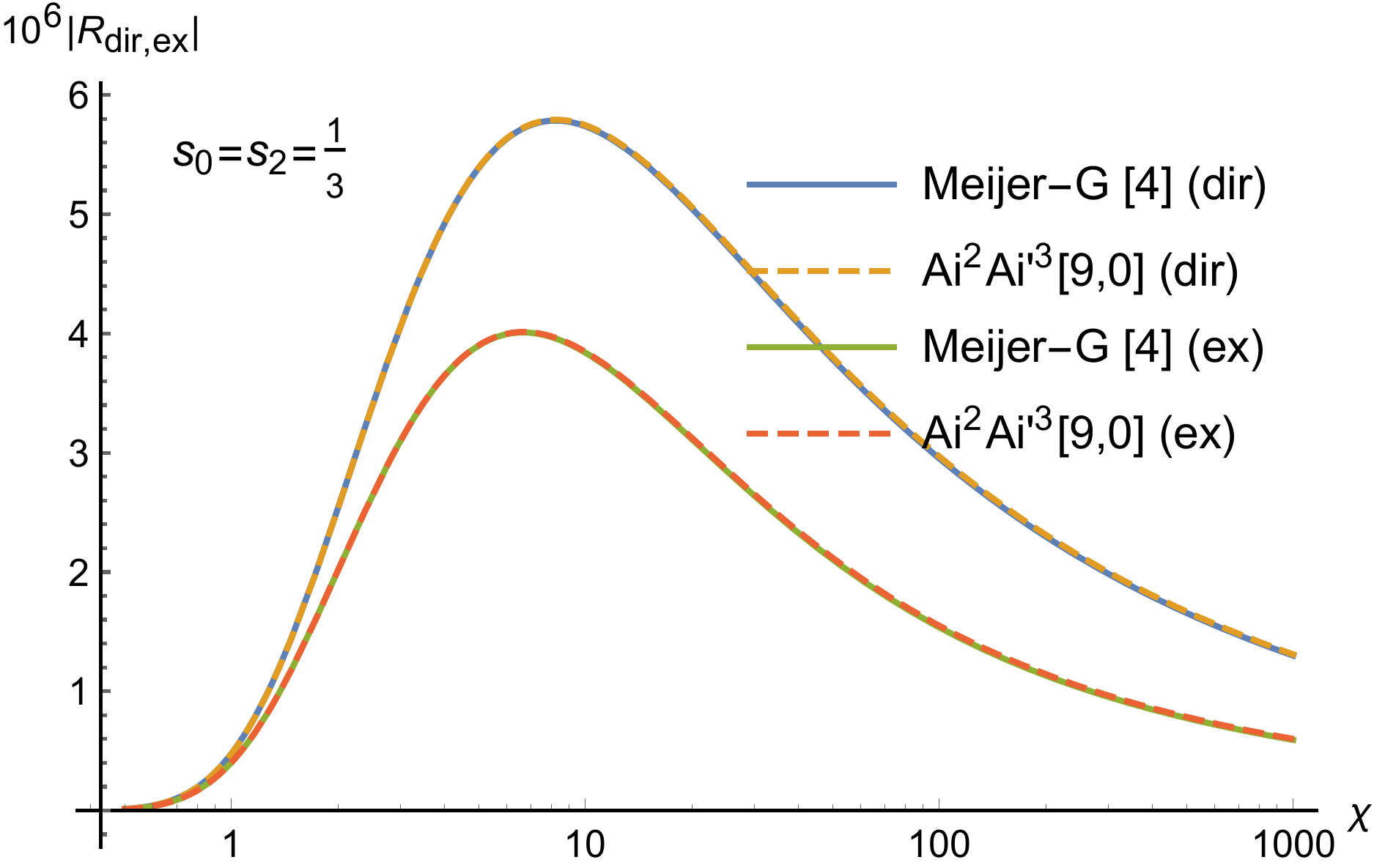}
\includegraphics[width=\linewidth]{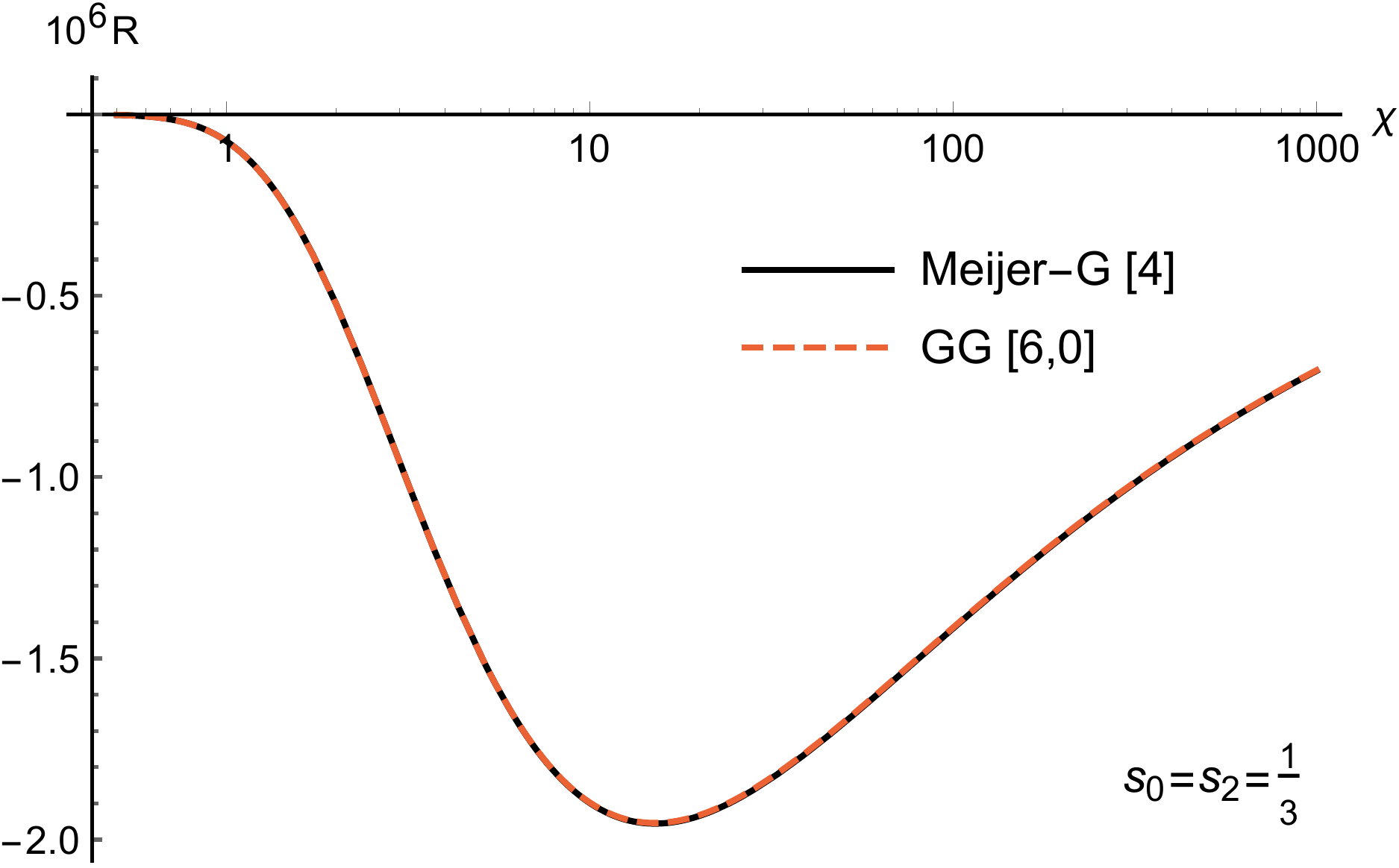}
\caption{First plot: exchange and minus direct part of the one-step, $R_{\rm ex}$ and $-R_{\rm dir}$. The large-$\chi$ lines are obtained with only large-$\chi$ expansion coefficients, resummed with the Meijer-G resummation~\eqref{sumofG}, and the $\text{Ai}^2\text{Ai}'^3$ lines are obtained with only the small-$\chi$ expansion, by resumming the first $9$ terms using~\eqref{AiryDerivResum}. Second plot: total one-step, $R_{\rm dir}+R_{\rm ex}$. Same notation as in Fig.~\ref{GGfig}.}
\label{DirExPhFig}
\end{figure}

As an example, we consider $s_0=s_2$, which is a saddle point (see Sec.~\ref{Saddle point approximations section}), and for definiteness $s_0=1/3$, which means that the final state particles share the initial longitudinal momentum equally. We showed in Sec.~\ref{Saddle point approximations section} that the direct and exchange parts of the one-step cancels to leading order for small $\chi$. However, as shown in Fig.~\ref{DirExPhFig}, for this example these terms do not cancel for larger $\chi$. This means that it makes more sense in this case to resum the direct and exchange parts separately, compared to the double-Compton example above where these terms cancel also for larger $\chi$. 

\begin{figure}
\includegraphics[width=\linewidth]{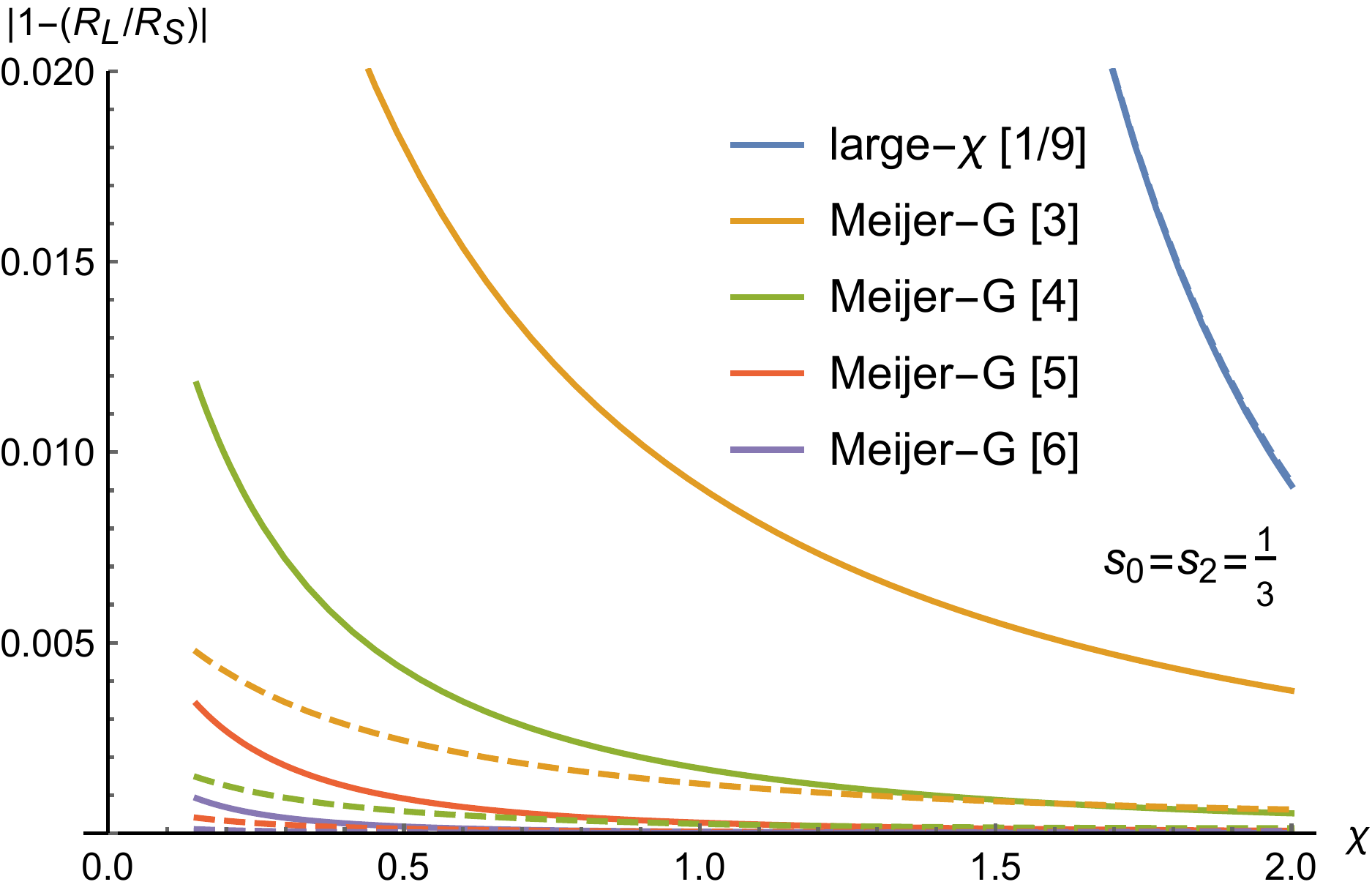}
\includegraphics[width=\linewidth]{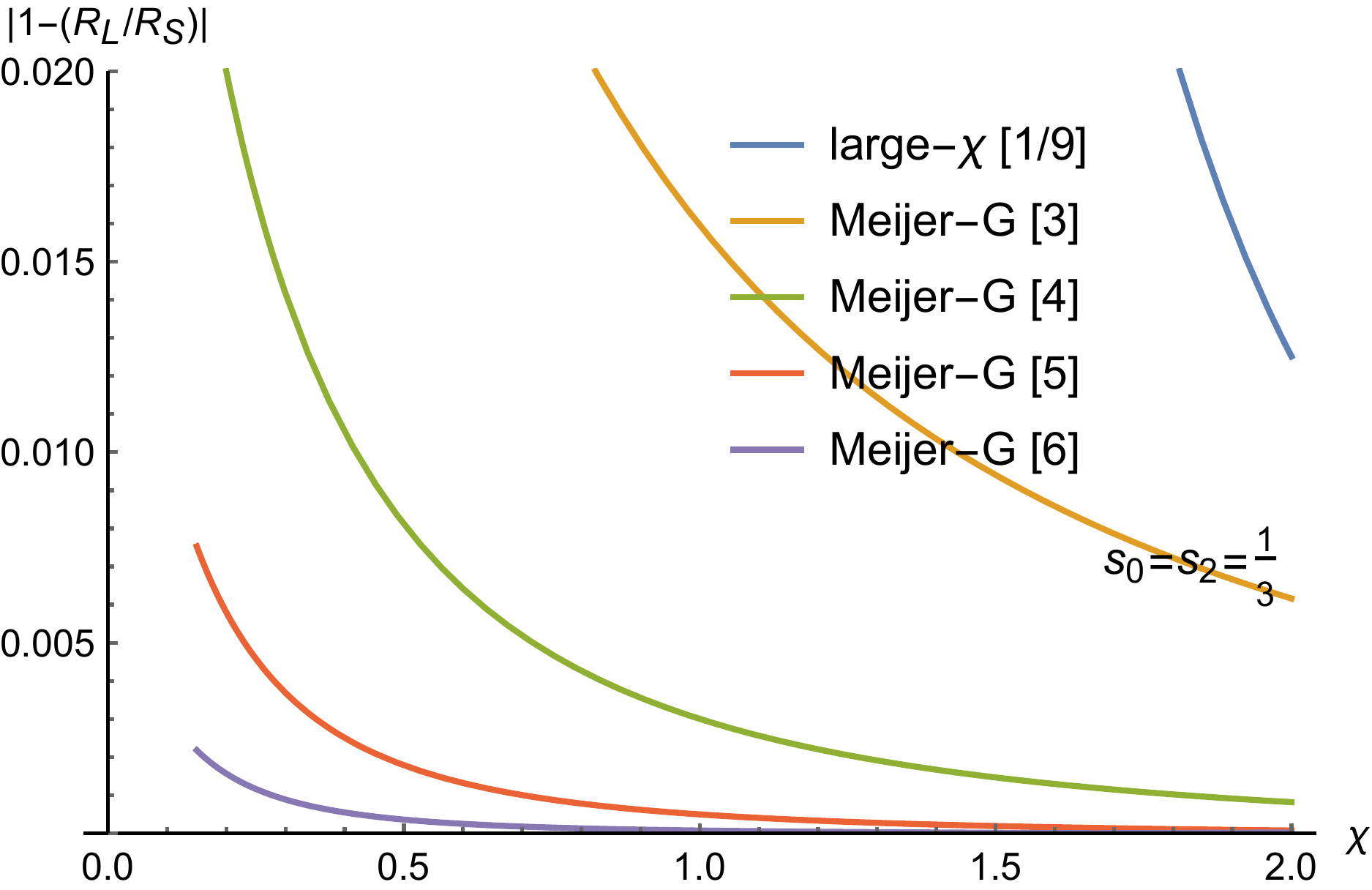}
\caption{The relative difference between large- and small-$\chi$ resummations, $R_{\rm L}$ and $R_{\rm S}$, for the direct (solid lines) and exchange parts (dashed lines) of the one-step in the first plot, and direct + exchange in the second plot. $R_{\rm S}$ has been obtained by resumming the small-$\chi$ expansion with Borel + Pad\'e, but Borel + conformal + Pad\'e or the confluent Hypergeometric resummation would give the same plots since the relative difference between these three is much smaller than $|1-R_{\rm L}/R_{\rm S}|$. The different lines correspond to different resummations of the large-$\chi$ expansion, with the same notation as in Fig.~\ref{MGresumRelErFig}. For the direct and exchange parts ``Meijer-G $[m]$'' corresponds to~\eqref{sumofG} with $n=-1,...,m-1$.}
\label{MGRelErPhFig}
\end{figure}

So, we first consider the resummation of the direct and exchange parts separately. Eventually one needs to sum the direct and exchange parts since it turns out that they are on the same order of magnitude and only their sum is gauge invariant. However, since the exchange part has a much more complicated integrand, it is useful to know whether or not there is some regime where it is negligible. To resum the small-$\chi$ expansions we can use the general methods described in Sec.~\ref{resumSmallSection}. The convergence properties of the large-$\chi$ expansion can be improved by off-diagonal Pad\'e approximants as in Sec.~\ref{resumLargeSection}. However, a much better resummation is achieved with the new Meijer-G resummation in Sec.~\ref{resumLargeSection}. The only difference is that the direct and exchange terms separately have small-$\chi$ expansions that starts with $(1/\sqrt{\chi})\exp(-2r/[3\chi])$ rather than $\sqrt{\chi}\exp(-2r/[3\chi])$ (cf.~\eqref{dirOneSmallExpansion}, \eqref{exOneSmallExpansion} and~\eqref{PoneSmallExpansion}), which can be taken into account simply by starting the sum in~\eqref{sumofG} with $n=-1$ instead of $n=0$. The improved precision is illustrated in Fig.~\ref{MGRelErPhFig}, where a much higher precision is obtained and even with fewer terms.

\begin{figure}
\includegraphics[width=\linewidth]{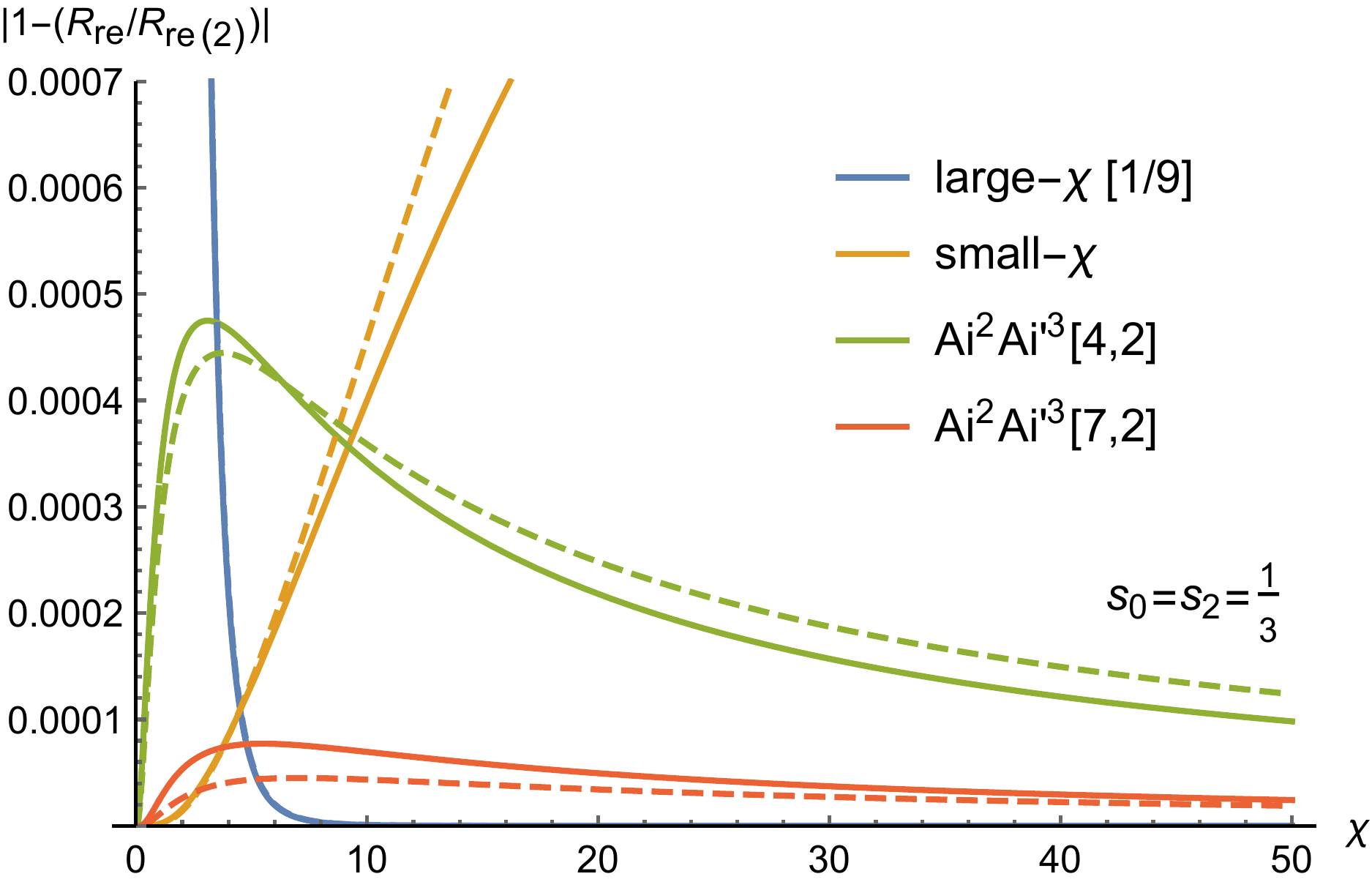}
\caption{Estimated relative error of resummations of small-, large-, or both small- and large-$\chi$ expansions. Solid and dashed lines correspond to the direct and exchange parts, respectively. The exact result has been approximated by~\eqref{sumofG} with $N=9$ for $\chi>1$ ($\chi>1/2$) for the direct (exchange) part and for smaller $\chi$ by the Borel + Pad\'e $[7/7]$ ($[5/5]$) resummation of the small-$\chi$ expansion. The the relative difference from the actual exact result is smaller than the scale of this plot (this can be estimated by comparing the two approximations at $\chi=1$ ($\chi=1/2$)).   
The small-$\chi$ lines have been obtained with the confluent hypergeometric resummation of the first $10$ terms, and, by testing different $b=n/100$ and comparing with the large-$\chi$ expansion at $\chi=10^3$, $b=74/100$ ($b=66/100$) for the direct (exchange) part. The large-$\chi$ lines are $[1/9]$ Pad\'e approximants of the large-$\chi$ expansion. The $\text{Ai}^2\text{Ai}'^3[m,n]$ lines correspond to~\eqref{AiryDerivResum} with the constants obtained by matching onto the first $(m,n)$ terms in the (small,large)-$\chi$ expansion.}
\label{uniformPhDirExFig}
\end{figure}

We can also use resummations similar to the ones in~\eqref{resumSmallLargeSection} to resum the small- and large-$\chi$ expansions simultaneously, which gives high precision for any value of $\chi$. However, we cannot use exactly the same resummations because the small-$\chi$ expansions have an overall factor of $1/\chi$ compared to the total one-step, while the large-$\chi$ expansions are still of the same type. Note that this means that one cannot simply apply the previous resummations to $\chi\mathbb{P}_{\rm dir}^{\rm one}$ and $\chi\mathbb{P}_{\rm ex}^{\rm one}$ because such an overall multiplication changes the large-$\chi$ expansion. So, instead of a resummation that only involves the Airy function, here we use both $\text{Ai}(z)$ and its derivative $\text{Ai}'(z)$. Let
\be
f_1(x)=\frac{2\sqrt{\pi}}{5^{1/6}}\text{Ai}\left[\frac{1}{(5x)^{2/3}}\right]
\ee      
and
\be
f_2(x)=-2\sqrt{\pi}5^{1/6}\text{Ai}'\left[\frac{1}{(5x)^{2/3}}\right] \;,
\ee
then $f_1^2(x)f_2^3(x)/x^{1/3}$ has the correct type of small- and large-$\chi$ expansions. So, as one possible resummation we take
\be\label{AiryDerivResum}
\begin{split}
\sum_{n=0}^N& c_n\frac{(1-v_n^2)^{1/6}}{x^{1/3}(1-w_n^2)^{1/6}}f_1\left[\frac{x}{1-v_n}\right]f_1\left[\frac{x}{1+v_n}\right] \\
\times&f_2\left[\frac{x}{1-w_n}\right]f_2\left[\frac{x}{1+w_n}\right]f_2(x) \;,
\end{split}
\ee 
where the $3N$ constants $c_n$, $v_n$ and $w_n$ can be obtained by matching with either the small-$\chi$ expansion or both the small- and large-$\chi$ expansions. For the example considered here, we find that that these constants are in general complex, as in the previous section, but this time a single solution for $c_n$, $v_n$ and $w_n$ can give resummations with a small imaginary part. However, this is not a problem because the complex conjugate $c_n^*$, $v_n^*$ and $w_n^*$ is also a solution, so by summing over both solutions one obtains real resummations. This is the same as simply taking the real part in cases where a single solution gives resummations with nonzero imaginary part, and we still only need $3N$ coefficients from the small/large-$\chi$ expansions.
As shown in Fig.~\ref{uniformPhDirExFig}, this Airy function approach allows us to obtain resummations with uniform precision, with a finite maximum relative error at some finite $\chi$. In general, if there are $N$ constants at a given order of some resummation and all of them are determined by the first $N$ coefficients of the large-$\chi$ expansion, then the result is usually more precise at large $\chi$ compared to a resummation where e.g. half of the constants are determined by the large-$\chi$ expansion and the rest from the small-$\chi$ expansion. And vice versa if all constants are determined by the small-$\chi$ expansion. However, using coefficients from both expansions (still with the number of constants, $N$, fixed) allows us to obtain resummations that is precise for arbitrary $\chi$, and it can be much faster to obtain e.g. the first 5 terms in the small- and 5 terms in the large-$\chi$ expansions rather than 10 terms in the small-$\chi$ expansion.              

We have now showed how the direct and exchange parts can be resummed separately. From this we see that the exchange term is on the same order of magnitude as the direct term also for large $\chi$, but they do not cancel in this case. This means that we can simply add the two separate resummations without losing precision. However, having seen that they are on the same order of magnitude and recalling that only their sum is gauge invariant, it is also natural to sum the two terms from the start and construct resummations of their sum. To resum the small-$\chi$ expansion we can for example use the $GG$ resummation in~\eqref{newQuadResum}. As shown in Fig.~\ref{DirExPhFig}, this approach allows us to obtain good precision up to large $\chi$ with relatively few terms. To resum the large-$\chi$ expansion we can use the $G$ resummation in~\eqref{sumofG}. The precision of this resummation is shown in Fig.~\ref{MGRelErPhFig}.

\begin{figure}
\includegraphics[width=\linewidth]{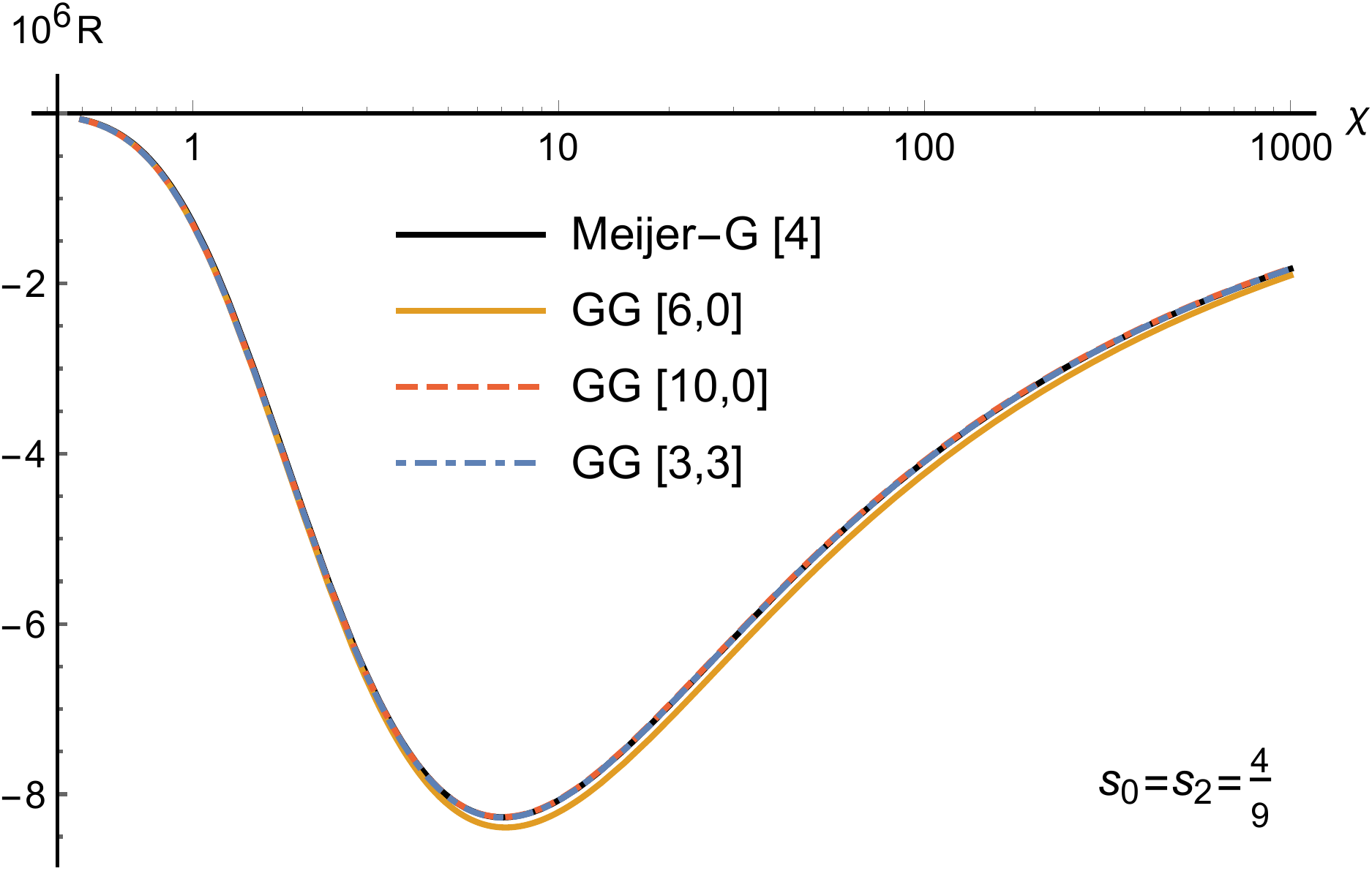}
\caption{Same notation as in Fig.~\ref{DirExPhFig}, different momentum.}
\label{DirExPhFig2}
\end{figure}

\begin{figure}
\includegraphics[width=\linewidth]{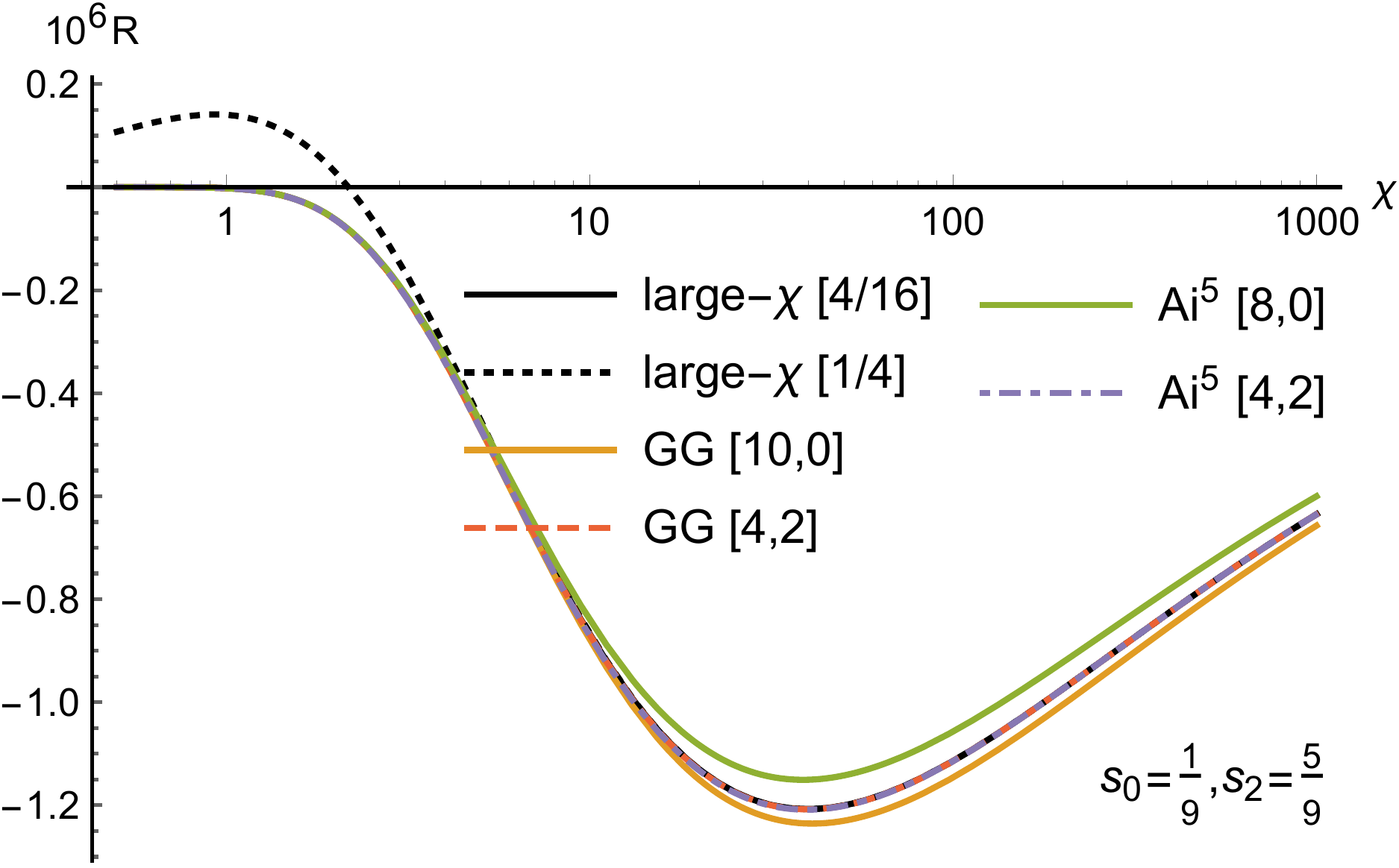}
\caption{Same notation as in Fig.~\ref{DirExPhFig}. $\text{Ai}^5$ is the resummation in~\eqref{Ai5ReVers1}. The large-$\chi$ $[4/16]$ is indistinguishable from the exact result on this scale, but diverges from it at small $\chi$.}
\label{DirExPhFig1d9a5d9}
\end{figure}

\begin{figure}
\includegraphics[width=\linewidth]{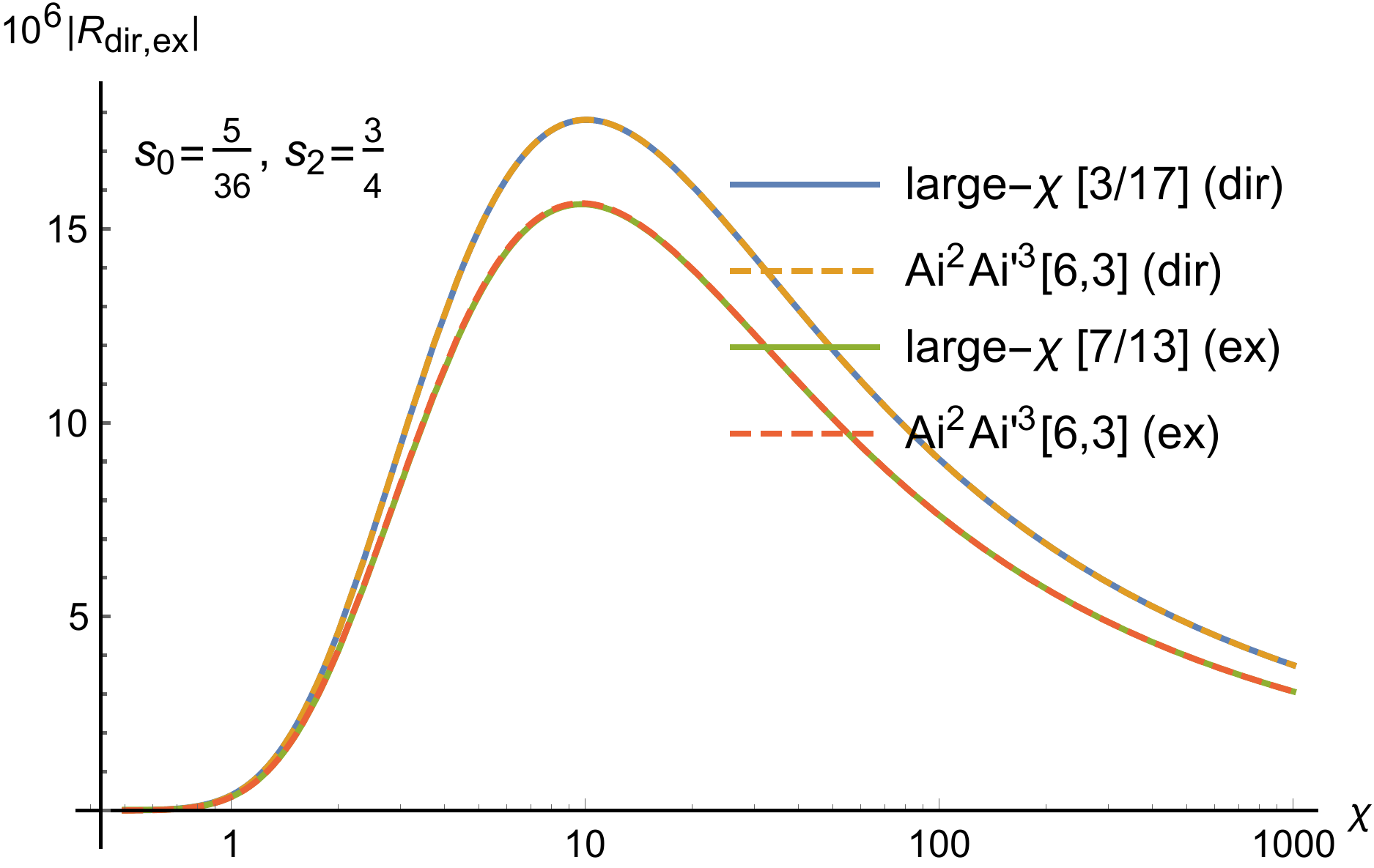}
\includegraphics[width=\linewidth]{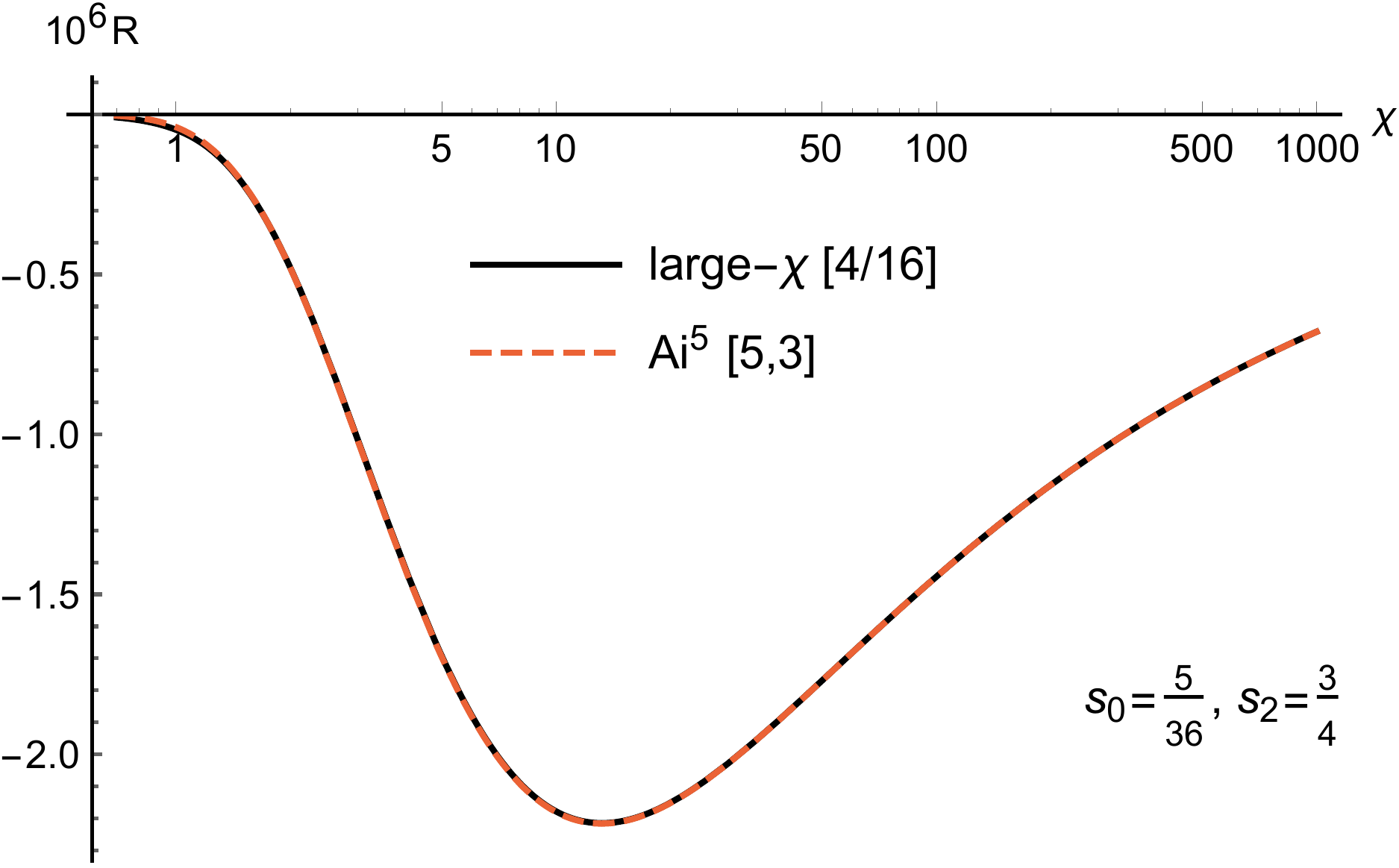}
\caption{Same notation as in Fig.~\ref{DirExPhFig}, different momentum.}
\label{DirExPhFig5d36a3d4}
\end{figure}

We have used $s_0=s_2=1/3$ as an example. The same methods can of course be used for other points in the spectrum. The only difference is how many terms from the expansions one needs. In Fig.~\ref{DirExPhFig2} we consider $s_0=s_2=4/9$, where the final-state photon has lower momentum compared to the pair. We find that $-\mathbb{P}_{\rm dir}^{\rm one}$ is roughly a factor of $2$ larger than $\mathbb{P}_{\rm ex}^{\rm one}$ for large $\chi$, so in this example too these two terms do not cancel each other.
We see that we again can obtain a large overlap between the resummations that only use either the small- or the large-$\chi$-expansion coefficients. In the previous example, Fig.~\ref{DirExPhFig}, we found a very good precision at large $\chi$ by resumming just the first $6$ terms in the small-$\chi$ expansion with the resummation in~\eqref{newQuadResum}. This time, the same order of resummation gives a small, but noticeable error. This can be fixed by using $10$ terms from the small-$\chi$ expansion. However, by using coefficients from both the small- and the large-$\chi$ expansions, we can find a similar agreement with only $3$ terms from each expansion.   

In some cases it can be challenging to obtain a high precision at large $\chi$ if one only has access to and only uses $\sim10$ of the first coefficients in the small-$\chi$ resummation. Fig.~\ref{DirExPhFig1d9a5d9} and~\ref{DirExPhFig5d36a3d4} show two such examples. However, the resummations in Sec.~\ref{resumSmallLargeSection} allow us to fix this by using just a couple of coefficients from the large-$\chi$ expansion. In Fig.~\ref{DirExPhFig1d9a5d9} we obtain good precision for arbitrary $\chi$ by using just $(4,2)$ coefficients from the (small,large)-$\chi$ expansions.  
The momentum of the final photon is the same in Fig.~\ref{DirExPhFig} and~\ref{DirExPhFig1d9a5d9}, but in Fig.~\ref{DirExPhFig} the fermion momenta are at the saddle point, $s_0=s_2$, while in Fig.~\ref{DirExPhFig1d9a5d9} $s_0$ and $s_2$ differ by a factor of $5$. As expected, this means that there is more exponential suppression in the second example.

In these photon trident examples, the direct and exchange parts do not cancel at larger $\chi$ in the way that they do in the double Compton examples. However, even for these photon trident examples there is a partial cancellation because the direct and exchange parts are on the same order of magnitude but have opposite sign. We have also found that the cancellation increases if one keeps $s_0+s_2$ fixed but moves away from the saddle point $s_0=s_2$. This is illustrated in Fig.~\ref{DirExPhFig5d36a3d4}, where $\mathbb{P}_{\rm ex}^{\rm one}$ is much closer to $-\mathbb{P}_{\rm dir}^{\rm one}$ compared to the case in Fig.~\ref{DirExPhFig2}, where $-\mathbb{P}_{\rm dir}^{\rm one}\gtrsim1.8\mathbb{P}_{\rm ex}^{\rm one}$ for $\chi>10$ and $-\mathbb{P}_{\rm dir}^{\rm one}\sim2.2\mathbb{P}_{\rm ex}^{\rm one}$ as $\chi\to\infty$.      

In this paper we have focused on resumming the one-step term. We could use these resummations methods also for the two-step. However, the two-step can anyway be expressed in terms of Airy functions, so a Meijer-G approach would simply lead to the same, exact result.

\section{Conclusions}\label{conclusionsSection}

We have studied the photon trident process in plane-wave backgrounds. In contrast to the other two $\mathcal{O}(\alpha^2)$ processes with only one incoming particle, trident and double Compton, photon trident has not attracted much attention, and we are only aware of one previous paper~\cite{MorozovNarozhnyiPhTr}. However, as these $\mathcal{O}(\alpha^2)$ processes are the first steps in the formation of cascades, it is important to study all three $\mathcal{O}(\alpha^2)$ processes. We have already showed that several results for double Compton can be obtained by making certain replacements in our results for trident. This is especially useful for the exchange term, because it is in general difficult to calculate and so it is good to know that one can use the same methods to compute these exchange terms in trident and double Compton. In this paper we have shown that there is an even closer relation between double Compton and photon trident. All terms can be obtained by replacing the longitudinal momenta in the double-Compton expressions. We have shown this explicitly for the leading order in $\chi\ll1$.

This means that we can immediately obtain saddle-point approximations for photon trident from our corresponding results in~\cite{Dinu:2018efz} for double Compton by just replacing the longitudinal momenta. In particular, this means that the direct and exchange parts of the one-step cancel each other to leading order, not just for double Compton, but also for photon trident.
One reason to consider trident, double Compton and photon trident is to better delineate the region of parameter space where a two-step approximation works, and by extension where a corresponding ``N-step'' approximation works for cascades. The near cancellation between the direct and exchange parts of the one-step in double Compton and photon trident is thus important as it tells us that the two-step approximation is better than what one would have otherwise guessed based on the scaling of the two-step and one-step with respect to $a_0$.
In the double-Compton case, this near-cancellation continues up to large $\chi$ for large parts of the spectrum. Here we have seen that this does not in general happen for photon trident, but the direct and exchange terms anyway continues to be on the same order of magnitude.

The one-step terms can be challenging to calculate, especially the exchange part. Here we have shown how the small- and large-$\chi$ expansions can be resummed to obtain a good precision for large intervals of $\chi$ or even arbitrary $\chi$. The small-$\chi$ expansion is divergent. We have showed that this series can be resummed with Borel transformation, conformal maps and Pad\'e approximants, or with a new resummation~\cite{Alvarez:2017sza} based on a confluent hypergeometric function. 
The large-$\chi$ expansion seems to be convergent and does not need resummation for $\chi$ larger than some fixed value. However, with a finite number of terms, one can significantly extend the large-$\chi$ expansion by suitable resummations. A first improvement can be obtained by (far from diagonal) Pad\'e approximants. However, the exact result has an exponential scaling at small $\chi$, so any Pad\'e approximant of the large-$\chi$ expansion eventually breaks down as $\chi$ decreases. For this reason we have developed new resummation methods, which have the same type of expansions as the exact result for both small and large $\chi$. These resummations can be expressed in terms of Meijer-G functions. We were inspired to look for such resummations by the Meijer-G resummation in~\cite{Mera:2018qte}. But in contrast to~\cite{Mera:2018qte}, we are dealing with a class of Meijer-G functions that have exponential rather than power-law scaling at large argument (small $\chi$). And our resummations involve sums of Meijer-G functions, rather than a single Meijer-G with increasing number of parameters. We have found that these new resummations work well for resummation of the one-step terms, including the exchange part. We expect this to be useful also for other processes and quantities in LCF, because the structure of the small- and large-$\chi$ expansions are largely determined by the exponential part of the lightfront-time integrands, which can in general, for all processes with a single particle in the initial state, be expressed in terms of Kibble's effective mass. 

One useful generalization of what has been studied here would be to consider quantities where the large-$\chi$ expansion has logarithmic terms, which is for example the case for the trident probability integrated over the longitudinal momenta. One can expect that it should be possible to treat also such cases with Meijer-G functions, as they have log terms for certain parameters.
It could also be useful to consider resummations of expansions in the longitudinal momenta of the final-state particles, as this might allow one to find results that work for both double Compton and photon trident (because they are related via replacements of the momenta).

\acknowledgements

G.~T. thanks Victor Dinu for discussions about photon trident.
G.~T. was supported by the Alexander von Humboldt foundation during the initial parts of this project.

\end{document}